\pdfoutput=1
\documentclass[a4paper,american,aps,citeautoscript,floatfix,pdftex,pra,showpacs,superscriptaddress,twoside,longbibliography,reprint,twocolumn]{revtex4-1}
\usepackage{amsmath,amssymb}
\usepackage{commath}
\usepackage{dcolumn}
\usepackage[T1]{fontenc}
\usepackage[utf8]{inputenc}
\usepackage{graphicx}
\usepackage{microtype}
\usepackage{textcomp}
\usepackage[textsize=footnotesize]{todonotes}
\usepackage{xspace}
\usepackage{hyperref, hypernat}
\usepackage[todos]{CMImacros}

\newlength{\figwidth}
\newcommand{\cfeldesy}{\affiliation{Center for Free-Electron Laser Science, Deutsches
      Elektronen-Synchrotron DESY, Notkestraße 85, 22607 Hamburg, Germany}}
\newcommand{\uhhphys}{\affiliation{Department of Physics, University of Hamburg, Luruper Chaussee
      149, 22761 Hamburg, Germany}}
\newcommand{\uhhcui}{\affiliation{The Hamburg Center for Ultrafast Imaging, University of Hamburg,
      Luruper Chaussee 149, 22761 Hamburg, Germany}}

\newcommand{\elettra}{\altaffiliation[Present address:~]{Elettra-Sincrotrone Trieste S.C.p.A.,
      34149, Basovizza, Trieste, Italy}}

\begin{document}
\title{Velocity-map imaging for emittance characterization \mbox{of multiphoton-emitted electrons
      from a gold surface}}
\author{Hong Ye}\cfeldesy\uhhphys
\author{Sebastian Trippel}\email[Email:~]{sebastian.trippel@cfel.de}\cfeldesy\uhhcui
\author{Michele Di Fraia}\elettra\cfeldesy\uhhcui
\author{Arya Fallahi}\cfeldesy
\author{Oliver D.~Mücke}\cfeldesy\uhhcui
\author{Franz X.~Kärtner}\cfeldesy\uhhphys\uhhcui
\author{Jochen Küpper}\cfeldesy\uhhphys\uhhcui
\date{\today}
\pacs{79.60.Bm, 41.20.Cv, 85.60.Ha, 79.20.Ws}

\begin{abstract}\noindent
   A velocity-map-imaging spectrometer is demonstrated to characterize the normalized transverse
   emittance of photoemitted electron bunches. The two-dimensional (2D) projected velocity
   distribution images of photoemitted electrons are recorded by the detection system and analyzed
   to obtain the normalized transverse emittance. With the presented distribution function of the
   electron photoemission angles a mathematical method is implemented to reconstruct the
   three-dimensional (3D) velocity distribution curve. As a first example, multiphoton emission from
   a planar Au surface is studied via irradiation at a glancing angle by intense 45~fs laser pulses
   at a central wavelength of 800~nm. The reconstructed energy distribution agrees very well with
   the Berglund-Spicer theory of photoemission. The normalized transverse emittance of the intrinsic
   electron bunch is characterized to be 0.52 and 0.05 $\pi\cdot\text{mm}\cdot\text{mrad}$ in $X$- and
   $Y$-directions, respectively.
\end{abstract}
\maketitle

\section{Introduction}
\label{sec:introduction}
Time-resolved imaging of both transient molecular structure and condensed phase dynamics with
picometer-femtosecond spatiotemporal resolution has recently become possible with the advent of
x-ray free-electron lasers (XFELs)~\cite{Chapman:NatPhys2:839, Barty:NatPhoton2:415,
   Young:Nature466:56, Barty:NatPhoton6:35, Rudek:NatPhoton6:858, Barty:ARPC64:415,
   Erk:Science345:288}. The high x-ray brilliance, coherence, and ultrashort pulse durations
available from these sources are the key properties~\cite{Huang:PRSTAB10:034801} that open up
unprecedented opportunities for new science. Therefore, precise control of the x-ray pulse
characteristics, including spectral coverage and temporal and spatial beam profiles are of utmost
importance for advanced applications. These parameters are directly influenced by the properties of
the electron bunch generating the x-ray pulses. Therefore, the accurate characterization of the
electron beam quality is indispensable for assessing available approaches in order to enable
improvements of the underlying electron beam technology. In addtion, high quality electron bunches
are instrumental in experiments where materials are studied using electron diffractive
imaging~\cite{Ihee:Science291:458, Siwick:Science302:1382, Gulde:Science345:200,
   Yang:PRL117:153002}.

The key measure in electron beam quality is electron beam emittance, \ie, the transverse phase-space
distribution of the generated electron bunches. To quantify electron beam emittance as a function of
photocathode composition and emission mechanisms, we demonstrate a velocity-map-imaging (VMI)
spectrometer that allows us to directly access the transverse momentum distribution of photoemitted
electrons, enabling the measurement of normalized transverse emittance from various cathodes.
Usually, emission mechanisms are classified as thermionic emission, photoemission, or tunneling
emission under extraordinarily high electric fields. More recently, nanostructured and plasmonic
photocathodes used with multiphoton or strong-field optical emission have been used as improved
electron sources~\cite{Kruger:Nature475:78, Herink:Nature483:193, Mustonen:APL99:103504,
   Putnam:NatPhys13:335, Tsujino:NatComm7:13976, Kaertner:NIMA829:24}. Both, the experimental
characterization and the theoretical description of the electron emittance from such cathodes is
highly important, which motivates the direct VMI measurements developed here.

As a first proof-of-principle example, we report on quantitative measurements of multiphoton
emission from a 400~nm thick Au thin film at room temperature, which was excited with 45-fs laser
pulses centered at 800~nm. These measurements additionally allowed us to benchmark the performance
of this new experimental setup. Quantum-yield-dependent measurements were performed by recording the
events of electrons impinging on the detector when varying the average laser power and the
polarization angle, respectively. These experimental results confirm that four-photon emission
occurs from the planar Au surface. In our experiments the 2D transverse velocity/momentum
distribution of photoemitted electrons was directly imaged onto the detector. An experimental 3D
energy distribution was reconstructed from the measured 2D VMI data using a mathematical algorithm
(\emph{vide infra}) and compared to the theoretically derived 3D-space energy distribution from the
Berglund-Spicer photoemission model~\cite{Berglund:PR136:A1030, Berglund:PR136:A1044,
   Krolikowski:PRB185:882, Krolikowski:PRB1:478}. The very good agreement of our experimental
results with the theoretical model demonstrates the applicability of VMI for the characterization of
the normalized transverse emittance of photoelectron emitters.

\section{Experimental Setup}
\label{sec:setup}
The velocity-mapping technique maps the velocity coordinates of particles onto a 2D detector
without, to first order, the influence of the spatial coordinates. To achieve this, a configuration
of electrostatic lenses, in the simplest case using three parallel electrodes, is employed to
spatially tailor the electric fields~\cite{Eppink:RSI68:3477, Chichinin:IRPC28:607,
   Stei:JCP138:214201}. The electric fields can be also used to image and magnify the spatial
coordinates suppressing the effect of velocity coordinates, which is then referred to as spatial-map
imaging (SMI)~\cite{Eppink:RSI68:3477}. The spectrometer demonstrated here aims to characterize the
electron emittance via characterizing the average spread of electron coordinates in
position-and-momentum phase space.

\begin{figure}
   \centering
   \includegraphics[width=\linewidth]{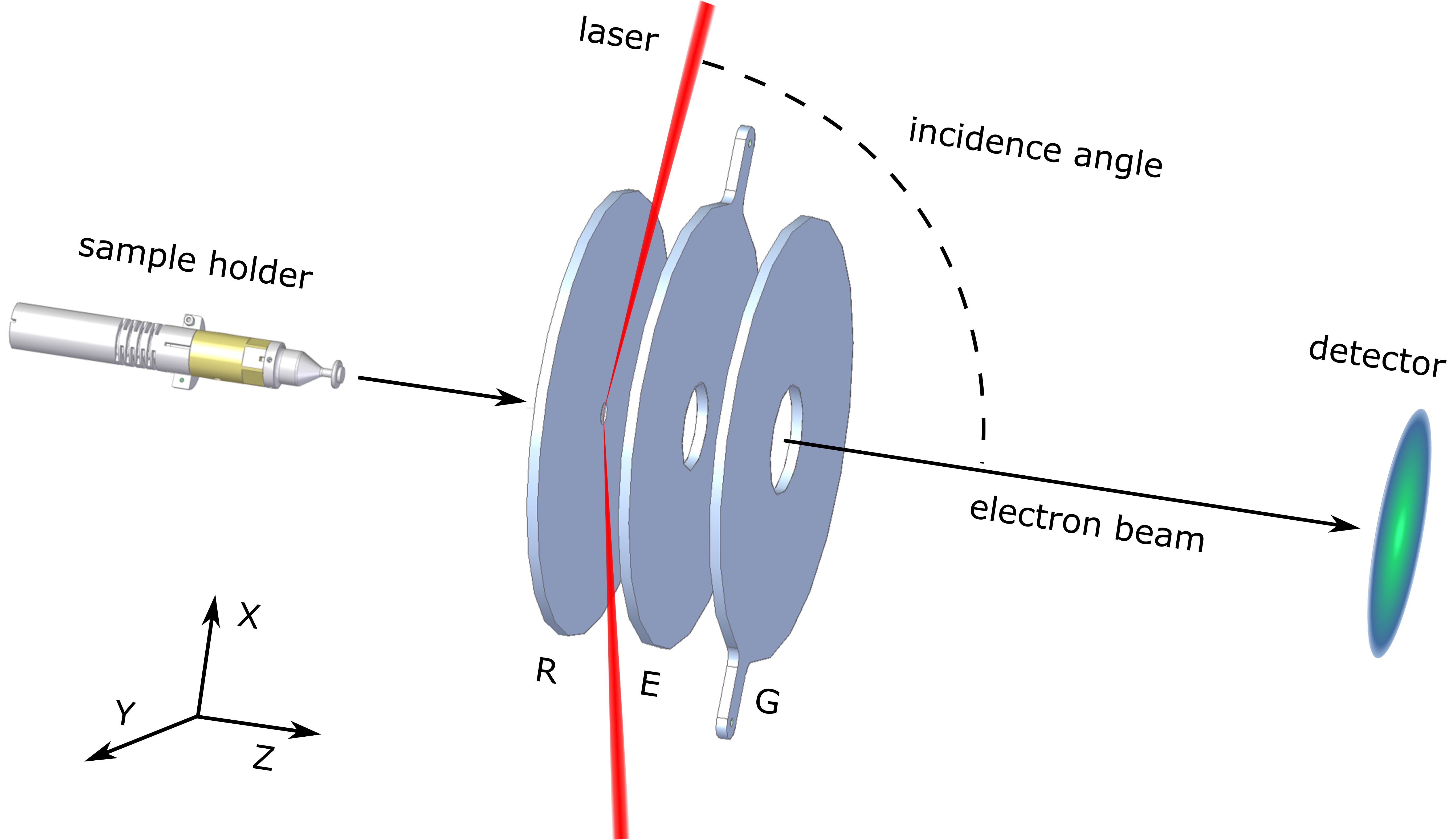}
   \caption{Schematic of the velocity-map-imaging (VMI) spectrometer consisting of three parallel
      electrodes, R: repeller, E: extractor, G: ground. The sample is mounted on the top of the
      holder, which can be retracted from this main chamber into a load-lock chamber.}
   \label{fig:VMI-setup}
\end{figure}
The schematic of the spectrometer is shown in \autoref{fig:VMI-setup}. The sample is mounted on the
top of the sample holder, which can be retracted into a load-lock chamber. The load lock is designed
for exchanging the sample without breaking the ultra-high vacuum (UHV) of the imaging system. When
performing the electrostatic imaging experiments, the sample holder is transferred into the main
chamber and brought in contact with the repeller plate to make sure they are at the same electric
potential. The main chamber, maintained at $10^{-9}~\text{mbar}$, contains a stack of three
cylindrically symmetric plates, labelled repeller~(R), extractor~(E) and ground~(G) electrodes in
\autoref{fig:VMI-setup}. They are arranged in parallel, separated by 15~mm, and, with applied
potentials, serve as the electrostatic lens. This is followed by a $\ordsim0.5$~m drift tube, which
ends with a detector assembly consisting of a double micro-channel plate~(MCP, Chevron
configuration), a phosphor screen~(P46), and a CMOS camera~(Optronis CL600$\times$2) for recording
images of the electron distributions. The full configuration is shielded against stray fields by a
\textmu-metal tube. A 800-nm 45-fs Ti:Sapphire laser amplifier with a 3-kHz repetition rate was used
to illuminate the sample at a glancing incidence angle of $\ordsim\degree{84}$, with a laser focal
intensity spot size of $\ordsim17\ordtimes160~\um^2$ root-mean-square~(RMS) on the sample. In our
experiments, single-shot electron-distribution images are read out at a repetition rate of 1~kHz,
limited by the camera-acquisition frame rate. The average number of electrons emitted per pulse is
on the order of one or less, thus, space charge effects, which were reported
before~\cite{Petite:PRB45:12210}, are excluded.

To calibrate and optimize the spectrometer field configuration for both SMI and VMI, a fixed
potential of -6~kV was applied to both the repeller plate and the sample holder while the
ground plate being grounded; see supplementary information for details. While scanning the extractor
voltage from -5.8~kV to -4.3~kV, we observed the focusing behavior of the electron bunch depending
on the extractor voltage~\cite{Mueller:JPB48:244001}. This behavior is revealed by the RMS of the
electron bunch size in $X$- and $Y$-directions on the detector shown in Fig.~S1 (supplementary
information). The SIMION~\cite{Simion:8.1} software is used to simulate the electric field
configuration and to calculate the electron trajectories from a 2D Gaussian source with
$\sigma_X=140~\um$ and $\sigma_Y=15~\um$, yielding an RMS behavior curve that fits the experimental
results. SMI is obtained at the minimum RMS size, \ie, at an extractor voltage of -5560~V,
corresponding to a magnification factor of 7.5. From the measured SMI data, the RMS size is analyzed
to be $\sigma_X=158~\um$ and $\sigma_Y=20~\um$, which is in good agreement with the simulated
electron bunch size and the laser focal spot size. The extractor voltage for VMI conditions is found
at -4790~V according to the SIMION simulations and the calibration factor of velocity-per-pixel is
$8014~\text{m/s/pixel}$ on the detector. The details of the simulations and experimental calibration
are described in the supplementary material. In order to minimize field distortions, the sample
front surface should be placed in the same plane as the repeller front surface. However, samples of
different thickness lead to a position offset with reference to the repeller front, which strongly
influences the field configuration. Therefore, the extractor voltage for operating in SMI and VMI
mode are optimized by voltage adjustments of $[50,-50]$~V and $[400,-200]$~V, respectively, to
correct for a position offset of $[-0.5,0.5]$~mm. In this case, re-adjusting the potential right
after exchanging a sample is necessary, but quick (\emph{vide infra}).

\section{Experimental Results}
\label{sec:results}
\begin{figure}
   \centering
   \includegraphics[width=\linewidth]{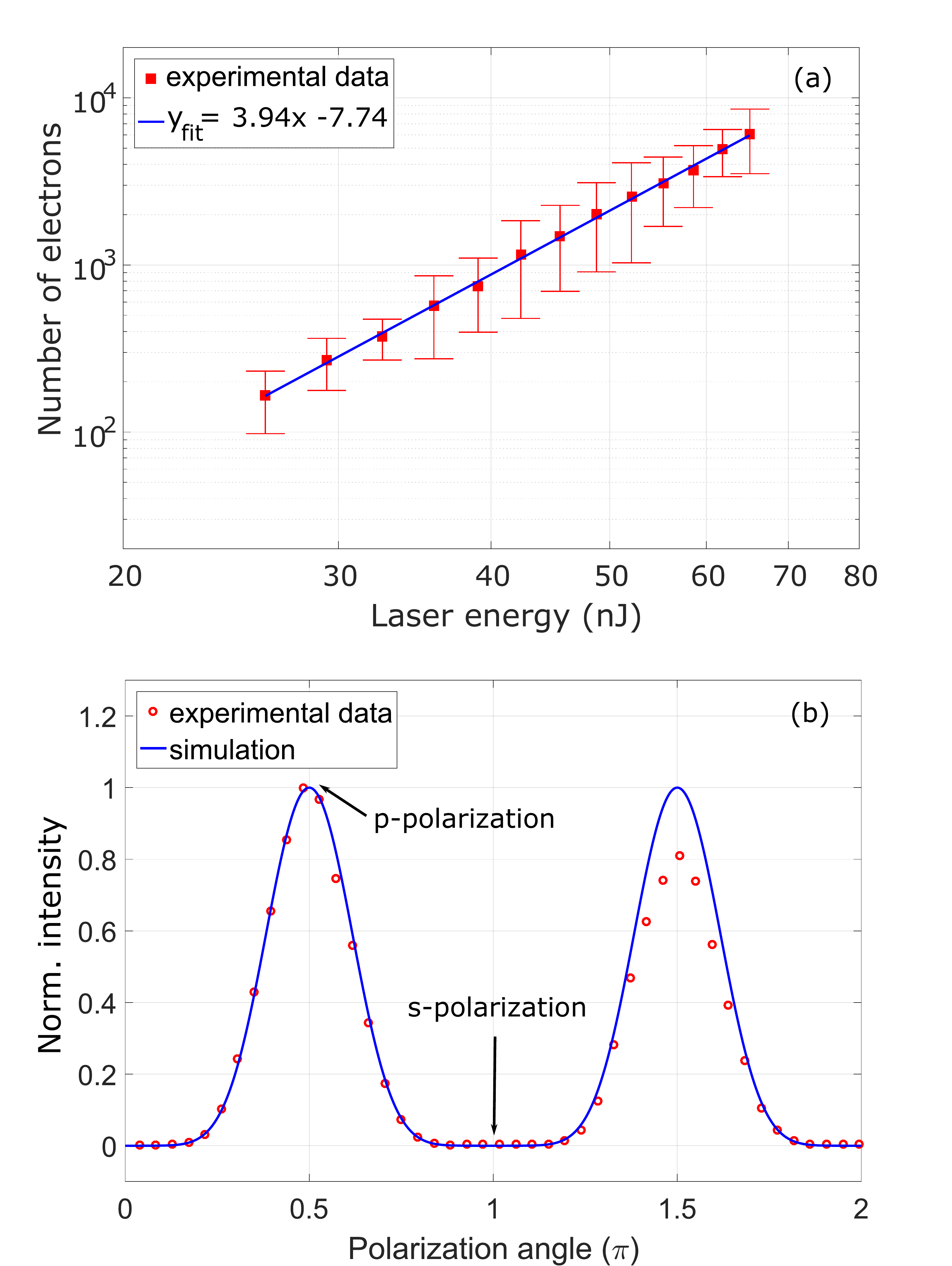}
   \caption{Counts of the photoemitted electrons as function of (a) laser average power and (b)
      laser polarization angle. The experimental data for polarization angles $>\pi$ is of reduced
      quality due to laser drifting, etc., within the errorbar plot in (a).}
   \label{fig:laser-dependence}
\end{figure}

\subautoref{fig:laser-dependence}{a} shows the photoemitted electron yield as a function of incident
laser energy on a logarithmic scale. The data, shown in red, were measured and averaged over four
measurement sequences and the error bars show the corresponding standard deviations of the
photoemitted electron counts due to laser fluctuations. The blue line reflects the results of a
linear regression analysis that yielded a slope of $c_x\approx3.94$, with a standard error of $0.04$
and a coefficient of determination $R^2\approx0.999$.

The Fowler-Dubridge model for the n-th order photoelectric current can be written in a generalized
form as~\cite{Bechtel:OPTCOMM13:56}
\begin{equation}
   J \propto A(1-R)^n \, I^n \, F\left(\frac{nh\nu-e\phi}{kT}\right)
\end{equation}
where $n$ is the number of photons, $h$ is the Planck constant, $A$ is the Richardson coefficient,
$R$ the reflection coefficient from the metal surface, $I$ the incident light intensity, $\phi$ the
metal work function, and $F(x)=\int_0^\infty\ln(1+e^{-(y+x)})\dif{y}$ the Fowler function.

The experimental data in \subautoref{fig:laser-dependence}{a} follow a power law with a slope of
$\ordsim4$, in agreement with a 4-photon emission process according to the nonlinear photoelectric
effect, which indicates that simultaneous absorption of 4 photons (photon energy 1.55~eV at 800~nm)
has to take place to overcome the metal work function $W$~\cite{Damascelli:PRB54:6031}, which is
reported as 5.31--5.47~eV for Au~\cite{Lide:CRCHandbookChemPhys:2003}. As shown in
\subautoref{fig:laser-dependence}{b}, varying the laser polarization angle, the photoemitted
electron intensity reaches a maximum when the laser is p-polarized (electric field normal to the
sample surface), and appears minimum when it is s-polarized. For multiphoton emission at a certain
incident light intensity, the electron yield mostly depends on the bulk absorption coefficient,
expressed as term $(1-R)^n$ in the Fowler-Dubridge model~\cite{Damascelli:PRB54:6031}. $R$ is
calculated by Fresnel equations with $n_1=1$ and $n_2=0.189+i4.71$~\cite{Polyanskiy:RID:2017} at an
incidence angle of \degree{84}. The plotted $(1-R)^4$ curve fits very well with the data, which
proves again the 4-th order multiphoton process.

\begin{figure}
   \centering
   \includegraphics[width=\linewidth]{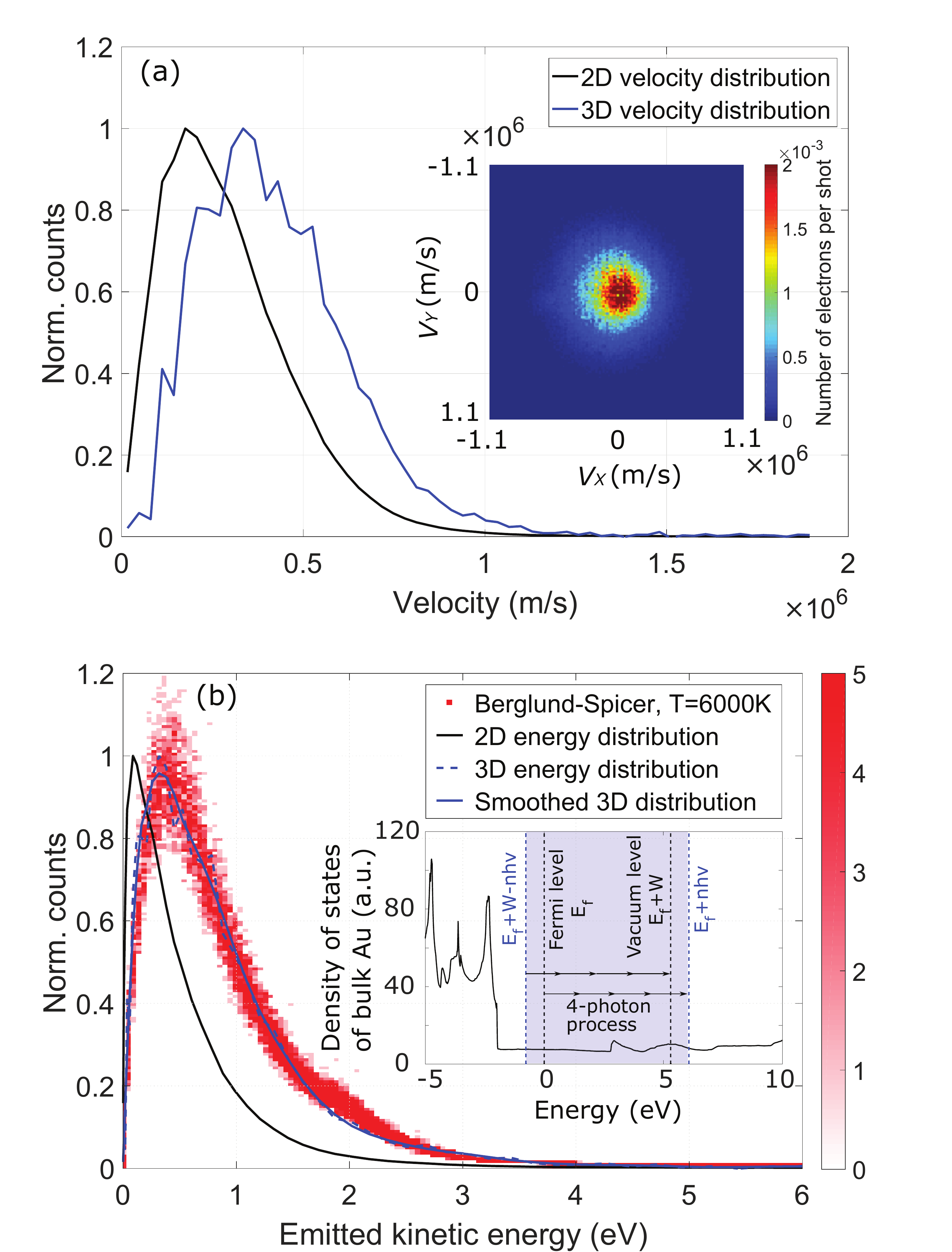}
   \caption{(a) Projected 2D (black curve) and reconstructed 3D (blue curve) radial velocity
      distribution of the measured velocity-map image that is shown in the inset. (b) Reconstructed
      kinetic-energy distribution and its simulation using the Berglund-Spicer model assuming an
      electron temperature of 6000~K. The colorbar represents the probabilities of photoelectron
      kinetic energies due to the photon-energy spectrum of the laser. The inset shows the density
      of states calculated for bulk Au, which is used in the Berglund-Spicer model simulation. The
      blue area depicts the four-photon-ionization range.}
   \label{fig:electron-distribution}
\end{figure}

A velocity-map image from a planar Au surface is shown in the inset of
\subautoref{fig:electron-distribution}{a}. The image was integrated over $6\ordtimes10^4$ laser
shots with an energy of $\ordsim50$~nJ, corresponding to a peak intensity of
$4\ordtimes10^{10}~\text{W/cm}^2$ on the cathode. Generally, in laser-induced multiphoton emission
the emitted electron velocity vectors exhibit cylindrical symmetry along the direction normal to the
sample surface. Therefore, the center of mass (COM) of the image is set as coordinate origin. The
corresponding angle-integrated radial velocity distribution of the projected electrons is plotted in
\subautoref{fig:electron-distribution}{a} as black line. To allow for comparison with the
theoretical model, the 3D velocity/energy distribution is required. Introducing a novel mathematical
method similar to the Onion Peeling algorithm~\cite{Dasch:AO31:1146}, we are able to reconstruct the
momentum/energy distribution when the angular distribution of emitted electrons is known.
Fortunately, for multiphoton emission, the intensity of photoemitted electrons at various angles
$\theta$ can be derived from the Berglund-Spicer model~\cite{Berglund:PR136:A1044} as
\begin{equation}
   I(\theta) \propto \aleph^2 \cos\theta\cdot\frac{1}{1+\alpha \,
      l(E)} \cdot \frac{1}{\sqrt{1-\aleph^2\cdot\sin^2\theta}}
\end{equation}
where $\alpha$ is the optical absorption coefficient, $l(E)$ is the electron-electron scattering
length for an electron of kinetic energy $E$, and $\aleph$ expresses the electron analogy of
refraction at the vacuum-metal boundary~\cite{Dowell:PRSTAB12:074201}. For a small $\aleph$ (our
case, $\aleph=0.275$), \ie, an incident photon energy $nh\nu$ comparable to the work function $W$,
the equation can be simplified to $I(\theta)\propto\cos\theta$~\cite{Poole:JESRP1:371,
   Pei:JJAP41:L52}. Therefore, the 3D velocity distribution can be reconstructed as is described in
detail in the supplementary information.

The reconstructed velocity distribution is plotted as blue line in
\subautoref{fig:electron-distribution}{a}, and the smoothed energy distribution shown in
\subautoref{fig:electron-distribution}{b}. The energy distribution of the emitted electrons shows an
energy spread of $\ordsim1$~eV, which corresponds to the energy difference between a
four-1.55~eV-photon excitation and the Au work function of 5.31~eV.

\section{Discussion}
The Berglund-Spicer three-step model is employed as the analytic expression for the kinetic energy
distribution of the photoemitted electrons. As the model is derived for single-photon emission, it
is implied in our analysis that the electrons at an initial energy state $E_0$ absorb sufficient
number of photons instantaneously, rather than sequentially, to be pumped to a higher energy state
$E=E_0+nh\nu$. The kinetic energy distribution for single-photon
emission~\cite{Berglund:PR136:A1030} is adapted to multiphoton emission as
\begin{align}
  \hspace*{-3pt}
  N(E)\dif{E} \propto &\frac{K \, C(E) \, \alpha}{\alpha+1/l(E)} \dif{E}\nonumber\\
                      &\times\left[1+4\left(\frac{E-E_f}{nh\nu}-1+\ln\frac{nh\nu}{E-E_f}\right)\right]
  \label{eq:N(E)dE}
\end{align}
where $E_f$ is the Fermi energy of Au. $C(E)=0.5\ordtimes(1-\sqrt{W/E})$ for $E\geqslant{}W$ is a
semiclassical threshold function. $l(E)$ is the electron-electron scattering length, which is
proportional to $E^{-3/2}$. The absorption efficient $\alpha$ is calculated from the extinction
coefficient $k=4.71$ as $\alpha=4\pi{}k/\lambda$ and taken as a constant
$\alpha=7.7\ordtimes10^{5}~\text{cm}^{-1}$ independent of electron energy. $K$ is a correction
factor related to both $C(E)$ and $\alpha\,l(E)$, which is between 0.5 to 1. To evaluate
\autoref{eq:N(E)dE}, the probability of a photon carrying energy $h\nu$ is calculated from the
measured laser spectrum in the range from 760 to 850~nm. To overcome the barrier of 5.31~eV, an
electron is assumed to always absorb four photons ($n\equiv4$). Absorption of various photon
energies leads to slight difference of the quantum yield at a certain emitted kinetic energy as one
can see from \subautoref{fig:electron-distribution}{b}. The main consequence of absorbing photons
with various energies is the spectral/intensity broadening, which is illustrated by the color coding
in \subautoref{fig:electron-distribution}{b}, but with an essentially unchanged spectral shape. We
mention that \eqref{eq:N(E)dE} only includes the emitted electrons that experience none or one
electron-electron scattering process during transport to the metal-vacuum surface. Electron-electron
scattering is dominant over electron-phonon scattering and reshapes the energy distribution on a
fast timescale, \ie, during an ultrashort laser pulse.

The density of states~(DOS), \ie, the number of states available for electrons at a certain energy
level, is shown in the inset of \subautoref{fig:electron-distribution}{b}. During the photoemission
process, an energy state $E_0$ is first occupied by an electron, which is then excited to a higher
energy state $E$, which was empty. As fermions, electrons obey the Pauli exclusion principle. In
thermal equilibrium, the possibility of electrons to occupy an available energy state is given by
the Fermi-Dirac~(FD) distribution $f_\text{FD}$. However, excitation of a metal with ultrashort
strong laser pulses initially creates a nonequilibrium distribution that then thermalizes via
electron-electron scattering towards a Fermi-Dirac distribution. In gold, this thermalization occurs
on a timescale of hundreds of femtoseconds~\cite{Fann:PRL68:2834, Fann:PRB46:13592}. Subsequently,
the electrons cool down by dissipating energy into the lattice via electron-phonon scattering
occurring on a longer picosecond timescale. In the following discussion, where we employ the
Berglund-Spicer model in our analysis, we assume that the electronic system can be described by a
Fermi-Dirac distribution with quasi-equilibrium electron temperature $T_e$. Hence, the appropriate
densities of states and FD distributions are multiplied with the energy distribution as
$N(E)\text{d}E\;f_\text{FD}(E_0)\;\text{DOS}(E_0)\;(1-f_\text{FD}(E))\;\text{DOS}(E)$, resulting in
the spectrum shown in \subautoref{fig:electron-distribution}{b}.

The best fit with our reconstructed experimental energy distribution is obtained for an electron
temperature of 6000~K. This is comparable to previously observed electron temperatures of 7000~K in
surface-enhanced multiphoton emission from copper~\cite{Aeschlimann:JCP102:8606}. The high energy
tail of the spectrum indicates that very ``hot'' electrons are photoemitted by the femtosecond laser
pulse, consistent with the high excess energy deposited into the electronic system. For the energy
tail up to 4~eV, except for the high temperature, another process that might need to be taken into
account is above-threshold photoemission (ATP), \ie, the absorption of one (or more) extra photon,
occurring together with the four-photon process~\cite{Banfi:PRL94:037601}. Moreover, for our
experimental conditions, we can neglect tunnel ionization, which could result in high energy emitted
electrons. Since we estimate the absorbed peak intensity for the recorded image,
\subautoref{fig:electron-distribution}{a}, to be $\ordsim4\ordtimes10^{9}~\text{W/cm}^2$ taking into
account Fresnel losses. This implies a Keldysh parameter $\gamma=\sqrt{W/2U_p}\approx17\gg1$, which
is well in the multiphoton emission regime; here, $U_p\propto\lambda^2I$ is the ponderomotive energy
with laser wavelength $\lambda$ and intensity $I$.

Since both, the measured quantum yield and the momentum distribution, are in quantitative agreement
with the Fowler-Dubridge and Berglund-Spicer models, as one would expect from multiphoton emission
from a planar Au cathode, the VMI spectrometer has successfully been implemented as a tool to
characterize the photoemitted electrons from cathodes, especially to directly measure the transverse
momentum distribution. Assuming there is no correlation between the location of emission and the
transverse momentum, the RMS normalized emittance $\epsilon_n$ is defined as
\cite{Dowell:PRSTAB12:074201}
\begin{equation}
   \epsilon_{n_{\zeta}}=\frac{\sqrt{\langle\zeta^2\rangle\langle{p_{\zeta}}^2\rangle}}{mc}
   \text{,~with~} \zeta\in\{X,Y\}
\end{equation}
where $\langle\zeta^2\rangle$ is the spatial spread and $\langle{p_{\zeta}}^2\rangle$ is the
momentum spread of the electron bunch. From the velocity map image shown in the inset of
\subautoref{fig:electron-distribution}{a}, the RMS normalized emittance of the planar Au
photocathode irradiated by 45-fs 800-nm laser pulses with a focal spot size of $\sigma_X=161~\um$
and $\sigma_Y=17~\um$ is characterized to be
$\epsilon_{n_X}=0.52$~$\pi\cdot\text{mm}\cdot\text{mrad}$ and
$\epsilon_{n_Y}=0.05$~$\pi\cdot\text{mm}\cdot\text{mrad}$ in the $X$ and $Y$-directions,
respectively. To decrease the intrinsic normalized emittance, in principle one needs to decrease
either the emission area or the momentum spread. The former can be intuitively decreased by an
extremely tight focal spot size or sharp tip surface, which geometrically limits the emission area.
For reducing of the momentum/energy spread, choosing a proper material with appropriate work
function and irritated by a laser beam with matched photon energy, for example the photoemission of
Cu under 266-nm laser irradiation, is expected to help. Further reduction is expected when entering
the strong-field emission regime, where the electrons are considered to adiabatically tunnel through
the surface barrier with zero initial momentum and are then driven by the instantaneous optical
field~\cite{Corkum:PRL71:1994, Kruger:Nature475:78}. Under these conditions electrons are expected
to be emitted with a relatively small divergence angle and significantly lower transverse momentum
spread.

\section{Conclusions}
We demonstrated an electron spectrometer with VMI and SMI capabilities, which intuitively allows for
the measurement of the normalized transverse emittance of photocathodes. \ie, through the direct
observation of the transverse position and momentum distributions. We verified and benchmarked the
capabilities of the instrument in a proof-of-concept experiment, in which we characterized the
photoemitted electrons from a 400~nm thin Au film. For ultrashort femtosecond laser pulses with a
peak intensity lower than $10^{12}~\text{W/cm}^2$ at 800~nm central wavelength, which would
correspond to $\gamma=1$, multiphoton emission is shown to be the dominant contribution to the
entire electron current.

We intend to utilize this new setup for the emittance characterization of electron bunches
strong-field emitted from nanotips under optical field irradiation. Such devices should show
superior emittance~\cite{Herink:Nature483:193, Tsujino:NatComm7:13976}. Moreover, the small radius
of the sharp tips realize a field enhancement, which dramatically lowers the laser power required
for entering the strong-field regime and thus avoids damaging of the cathodes. Our ongoing work aims
at the characterization of electron emission from nanostructured array emitters, which are predicted
to provide high-current low-emittance coherent electron bunches in the strong-field emission regime.

\section{Acknowledgments}
We gratefully acknowledge helpful discussions with Jens S.\ Kienitz and Nele Müller, the expert
technical support of Thomas Tilp, and Nicolas Tancogne-Dejean for providing the computed
density-of-states data of bulk Au.

Besides DESY, this work has been supported by the excellence cluster ``The Hamburg Center for
Ultrafast Imaging -- Structure, Dynamics and Control of Matter at the Atomic Scale'' (CUI,
DFG-EXC1074), the priority program QUTIF (SPP1840 SOLSTICE) of the Deutsche Forschungsgemeinschaft,
the European Research Council under the European Union's Seventh Framework Programme (FP7/2007-2013)
through the Consolidator Grant COMOTION (ERC-Küpper-614507) and the Synergy Grant AXSIS
(ERC-Kaertner-609920), the Helmholtz Association ``Initiative and Networking Fund'', and the
accelerator on a chip program (ACHIP) funded by the Betty and Gordon Moore foundation.

\bibliography{string,cmi}

\begin{thebibliography}{41}%
\makeatletter
\providecommand \@ifxundefined [1]{%
 \@ifx{#1\undefined}
}%
\providecommand \@ifnum [1]{%
 \ifnum #1\expandafter \@firstoftwo
 \else \expandafter \@secondoftwo
 \fi
}%
\providecommand \@ifx [1]{%
 \ifx #1\expandafter \@firstoftwo
 \else \expandafter \@secondoftwo
 \fi
}%
\providecommand \natexlab [1]{#1}%
\providecommand \enquote  [1]{``#1''}%
\providecommand \bibnamefont  [1]{#1}%
\providecommand \bibfnamefont [1]{#1}%
\providecommand \citenamefont [1]{#1}%
\providecommand \href@noop [0]{\@secondoftwo}%
\providecommand \href [0]{\begingroup \@sanitize@url \@href}%
\providecommand \@href[1]{\@@startlink{#1}\@@href}%
\providecommand \@@href[1]{\endgroup#1\@@endlink}%
\providecommand \@sanitize@url [0]{\catcode `\\12\catcode `\$12\catcode
  `\&12\catcode `\#12\catcode `\^12\catcode `\_12\catcode `\%12\relax}%
\providecommand \@@startlink[1]{}%
\providecommand \@@endlink[0]{}%
\providecommand \url  [0]{\begingroup\@sanitize@url \@url }%
\providecommand \@url [1]{\endgroup\@href {#1}{\urlprefix }}%
\providecommand \urlprefix  [0]{URL }%
\providecommand \Eprint [0]{\href }%
\providecommand \doibase [0]{http://dx.doi.org/}%
\providecommand \selectlanguage [0]{\@gobble}%
\providecommand \bibinfo  [0]{\@secondoftwo}%
\providecommand \bibfield  [0]{\@secondoftwo}%
\providecommand \translation [1]{[#1]}%
\providecommand \BibitemOpen [0]{}%
\providecommand \bibitemStop [0]{}%
\providecommand \bibitemNoStop [0]{.\EOS\space}%
\providecommand \EOS [0]{\spacefactor3000\relax}%
\providecommand \BibitemShut  [1]{\csname bibitem#1\endcsname}%
\let\auto@bib@innerbib\@empty
\bibitem [{\citenamefont {Chapman}\ \emph {et~al.}(2006)\citenamefont
  {Chapman}, \citenamefont {Barty}, \citenamefont {Bogan}, \citenamefont
  {Boutet}, \citenamefont {Frank}, \citenamefont {Hau-Riege}, \citenamefont
  {Marchesini}, \citenamefont {Woods}, \citenamefont {Bajt}, \citenamefont
  {Benner}, \citenamefont {A.}, \citenamefont {Pl{\"o}njes}, \citenamefont
  {Kuhlmann}, \citenamefont {Treusch}, \citenamefont {D{\"u}sterer},
  \citenamefont {Tschentscher}, \citenamefont {Schneider}, \citenamefont
  {Spiller}, \citenamefont {M{\"o}ller}, \citenamefont {Bostedt}, \citenamefont
  {Hoener}, \citenamefont {Shapiro}, \citenamefont {Hodgson}, \citenamefont
  {van~der Spoel}, \citenamefont {Burmeister}, \citenamefont {Bergh},
  \citenamefont {Caleman}, \citenamefont {Huldt}, \citenamefont {Seibert},
  \citenamefont {Maia}, \citenamefont {Lee}, \citenamefont {Sz{\"o}ke},
  \citenamefont {Timneanu},\ and\ \citenamefont
  {Hajdu}}]{Chapman:NatPhys2:839}%
  \BibitemOpen
  \bibfield  {author} {\bibinfo {author} {\bibfnamefont {H.~N.}\ \bibnamefont
  {Chapman}}, \bibinfo {author} {\bibfnamefont {A.}~\bibnamefont {Barty}},
  \bibinfo {author} {\bibfnamefont {M.~J.}\ \bibnamefont {Bogan}}, \bibinfo
  {author} {\bibfnamefont {S.}~\bibnamefont {Boutet}}, \bibinfo {author}
  {\bibfnamefont {S.}~\bibnamefont {Frank}}, \bibinfo {author} {\bibfnamefont
  {S.~P.}\ \bibnamefont {Hau-Riege}}, \bibinfo {author} {\bibfnamefont
  {S.}~\bibnamefont {Marchesini}}, \bibinfo {author} {\bibfnamefont {B.~W.}\
  \bibnamefont {Woods}}, \bibinfo {author} {\bibfnamefont {S.}~\bibnamefont
  {Bajt}}, \bibinfo {author} {\bibfnamefont {W.~H.}\ \bibnamefont {Benner}},
  \bibinfo {author} {\bibfnamefont {London~W.}\ \bibnamefont {A.}}, \bibinfo
  {author} {\bibfnamefont {E.}~\bibnamefont {Pl{\"o}njes}}, \bibinfo {author}
  {\bibfnamefont {M.}~\bibnamefont {Kuhlmann}}, \bibinfo {author}
  {\bibfnamefont {R.}~\bibnamefont {Treusch}}, \bibinfo {author} {\bibfnamefont
  {S.}~\bibnamefont {D{\"u}sterer}}, \bibinfo {author} {\bibfnamefont
  {T.}~\bibnamefont {Tschentscher}}, \bibinfo {author} {\bibfnamefont {J.~R.}\
  \bibnamefont {Schneider}}, \bibinfo {author} {\bibfnamefont {E.}~\bibnamefont
  {Spiller}}, \bibinfo {author} {\bibfnamefont {T.}~\bibnamefont {M{\"o}ller}},
  \bibinfo {author} {\bibfnamefont {C.}~\bibnamefont {Bostedt}}, \bibinfo
  {author} {\bibfnamefont {M.}~\bibnamefont {Hoener}}, \bibinfo {author}
  {\bibfnamefont {D.~A.}\ \bibnamefont {Shapiro}}, \bibinfo {author}
  {\bibfnamefont {K.~O.}\ \bibnamefont {Hodgson}}, \bibinfo {author}
  {\bibfnamefont {D.}~\bibnamefont {van~der Spoel}}, \bibinfo {author}
  {\bibfnamefont {F.}~\bibnamefont {Burmeister}}, \bibinfo {author}
  {\bibfnamefont {M.}~\bibnamefont {Bergh}}, \bibinfo {author} {\bibfnamefont
  {C.}~\bibnamefont {Caleman}}, \bibinfo {author} {\bibfnamefont
  {G.}~\bibnamefont {Huldt}}, \bibinfo {author} {\bibfnamefont {M.~M.}\
  \bibnamefont {Seibert}}, \bibinfo {author} {\bibfnamefont {F.~R. N.~C.}\
  \bibnamefont {Maia}}, \bibinfo {author} {\bibfnamefont {R.~W.}\ \bibnamefont
  {Lee}}, \bibinfo {author} {\bibfnamefont {A.}~\bibnamefont {Sz{\"o}ke}},
  \bibinfo {author} {\bibfnamefont {N.}~\bibnamefont {Timneanu}}, \ and\
  \bibinfo {author} {\bibfnamefont {J.}~\bibnamefont {Hajdu}},\ }\bibfield
  {title} {\enquote {\bibinfo {title} {Femtosecond diffractive imaging with a
  soft-x-ray free-electron laser},}\ }\href {\doibase 10.1038/nphys461}
  {\bibfield  {journal} {\bibinfo  {journal} {Nat. Phys.}\ }\textbf {\bibinfo
  {volume} {2}},\ \bibinfo {pages} {839--843} (\bibinfo {year}
  {2006})}\BibitemShut {NoStop}%
\bibitem [{\citenamefont {Barty}\ \emph {et~al.}(2008)\citenamefont {Barty},
  \citenamefont {Boutet}, \citenamefont {Bogan}, \citenamefont {Hau-Riege},
  \citenamefont {Marchesini}, \citenamefont {Sokolowski-Tinten}, \citenamefont
  {Stojanovic}, \citenamefont {Tobey}, \citenamefont {Ehrke}, \citenamefont
  {Cavalleri}, \citenamefont {D{\"u}sterer}, \citenamefont {Frank},
  \citenamefont {Bajt}, \citenamefont {Woods}, \citenamefont {Seibert},
  \citenamefont {Hajdu}, \citenamefont {Treusch},\ and\ \citenamefont
  {Chapman}}]{Barty:NatPhoton2:415}%
  \BibitemOpen
  \bibfield  {author} {\bibinfo {author} {\bibfnamefont {Anton}\ \bibnamefont
  {Barty}}, \bibinfo {author} {\bibfnamefont {Sebastien}\ \bibnamefont
  {Boutet}}, \bibinfo {author} {\bibfnamefont {Michael~J.}\ \bibnamefont
  {Bogan}}, \bibinfo {author} {\bibfnamefont {Stefan}\ \bibnamefont
  {Hau-Riege}}, \bibinfo {author} {\bibfnamefont {Stefano}\ \bibnamefont
  {Marchesini}}, \bibinfo {author} {\bibfnamefont {Klaus}\ \bibnamefont
  {Sokolowski-Tinten}}, \bibinfo {author} {\bibfnamefont {Nikola}\ \bibnamefont
  {Stojanovic}}, \bibinfo {author} {\bibfnamefont {Ra'anan}\ \bibnamefont
  {Tobey}}, \bibinfo {author} {\bibfnamefont {Henri}\ \bibnamefont {Ehrke}},
  \bibinfo {author} {\bibfnamefont {Andrea}\ \bibnamefont {Cavalleri}},
  \bibinfo {author} {\bibfnamefont {Stefan}\ \bibnamefont {D{\"u}sterer}},
  \bibinfo {author} {\bibfnamefont {Matthias}\ \bibnamefont {Frank}}, \bibinfo
  {author} {\bibfnamefont {S\v{a}sa}\ \bibnamefont {Bajt}}, \bibinfo {author}
  {\bibfnamefont {Bruce~W.}\ \bibnamefont {Woods}}, \bibinfo {author}
  {\bibfnamefont {M.~Marvin}\ \bibnamefont {Seibert}}, \bibinfo {author}
  {\bibfnamefont {Janos}\ \bibnamefont {Hajdu}}, \bibinfo {author}
  {\bibfnamefont {Rolf}\ \bibnamefont {Treusch}}, \ and\ \bibinfo {author}
  {\bibfnamefont {Henry~N.}\ \bibnamefont {Chapman}},\ }\bibfield  {title}
  {\enquote {\bibinfo {title} {Ultrafast single-shot diffraction imaging of
  nanoscale dynamics},}\ }\href {\doibase 10.1038/nphoton.2008.128} {\bibfield
  {journal} {\bibinfo  {journal} {Nat. Photon.}\ }\textbf {\bibinfo {volume}
  {2}},\ \bibinfo {pages} {415} (\bibinfo {year} {2008})}\BibitemShut {NoStop}%
\bibitem [{\citenamefont {Young}\ \emph {et~al.}(2010)\citenamefont {Young},
  \citenamefont {Kanter}, \citenamefont {Kraessig}, \citenamefont {Li},
  \citenamefont {March}, \citenamefont {Pratt}, \citenamefont {Santra},
  \citenamefont {Southworth}, \citenamefont {Rohringer}, \citenamefont
  {DiMauro}, \citenamefont {Doumy}, \citenamefont {Roedig}, \citenamefont
  {Berrah}, \citenamefont {Fang}, \citenamefont {Hoener}, \citenamefont
  {Bucksbaum}, \citenamefont {Cryan}, \citenamefont {Ghimire}, \citenamefont
  {Glownia}, \citenamefont {Reis}, \citenamefont {Bozek}, \citenamefont
  {Bostedt},\ and\ \citenamefont {Messerschmidt}}]{Young:Nature466:56}%
  \BibitemOpen
  \bibfield  {author} {\bibinfo {author} {\bibfnamefont {L}~\bibnamefont
  {Young}}, \bibinfo {author} {\bibfnamefont {E~P}\ \bibnamefont {Kanter}},
  \bibinfo {author} {\bibfnamefont {Bertold}\ \bibnamefont {Kraessig}},
  \bibinfo {author} {\bibfnamefont {Y}~\bibnamefont {Li}}, \bibinfo {author}
  {\bibfnamefont {A~M}\ \bibnamefont {March}}, \bibinfo {author} {\bibfnamefont
  {S~T}\ \bibnamefont {Pratt}}, \bibinfo {author} {\bibfnamefont
  {R}~\bibnamefont {Santra}}, \bibinfo {author} {\bibfnamefont {S.~H.}\
  \bibnamefont {Southworth}}, \bibinfo {author} {\bibfnamefont {N}~\bibnamefont
  {Rohringer}}, \bibinfo {author} {\bibfnamefont {L~F}\ \bibnamefont
  {DiMauro}}, \bibinfo {author} {\bibfnamefont {Gilles}\ \bibnamefont {Doumy}},
  \bibinfo {author} {\bibfnamefont {C~A}\ \bibnamefont {Roedig}}, \bibinfo
  {author} {\bibfnamefont {N}~\bibnamefont {Berrah}}, \bibinfo {author}
  {\bibfnamefont {L}~\bibnamefont {Fang}}, \bibinfo {author} {\bibfnamefont
  {M}~\bibnamefont {Hoener}}, \bibinfo {author} {\bibfnamefont {P~H}\
  \bibnamefont {Bucksbaum}}, \bibinfo {author} {\bibfnamefont {J~P}\
  \bibnamefont {Cryan}}, \bibinfo {author} {\bibfnamefont {S}~\bibnamefont
  {Ghimire}}, \bibinfo {author} {\bibfnamefont {J~M}\ \bibnamefont {Glownia}},
  \bibinfo {author} {\bibfnamefont {D~A}\ \bibnamefont {Reis}}, \bibinfo
  {author} {\bibfnamefont {J~D}\ \bibnamefont {Bozek}}, \bibinfo {author}
  {\bibfnamefont {C}~\bibnamefont {Bostedt}}, \ and\ \bibinfo {author}
  {\bibfnamefont {M}~\bibnamefont {Messerschmidt}},\ }\bibfield  {title}
  {\enquote {\bibinfo {title} {Femtosecond electronic response of atoms to
  ultra-intense x-rays},}\ }\href {\doibase 10.1038/nature09177} {\bibfield
  {journal} {\bibinfo  {journal} {Nature}\ }\textbf {\bibinfo {volume} {466}},\
  \bibinfo {pages} {56} (\bibinfo {year} {2010})}\BibitemShut {NoStop}%
\bibitem [{\citenamefont {Barty}\ \emph {et~al.}(2012)\citenamefont {Barty},
  \citenamefont {Caleman}, \citenamefont {Aquila}, \citenamefont {Timneanu},
  \citenamefont {Lomb}, \citenamefont {White}, \citenamefont {Andreasson},
  \citenamefont {Arnlund}, \citenamefont {Bajt}, \citenamefont {Barends},
  \citenamefont {Barthelmess}, \citenamefont {Bogan}, \citenamefont {Bostedt},
  \citenamefont {Bozek}, \citenamefont {Coffee}, \citenamefont {Coppola},
  \citenamefont {Davidsson}, \citenamefont {Deponte}, \citenamefont {Doak},
  \citenamefont {Ekeberg}, \citenamefont {Elser}, \citenamefont {Epp},
  \citenamefont {Erk}, \citenamefont {Fleckenstein}, \citenamefont {Foucar},
  \citenamefont {Fromme}, \citenamefont {Graafsma}, \citenamefont {Gumprecht},
  \citenamefont {Hajdu}, \citenamefont {Hampton}, \citenamefont {Hartmann},
  \citenamefont {Hartmann}, \citenamefont {Hauser}, \citenamefont {Hirsemann},
  \citenamefont {Holl}, \citenamefont {Hunter}, \citenamefont {Johansson},
  \citenamefont {Kassemeyer}, \citenamefont {Kimmel}, \citenamefont {Kirian},
  \citenamefont {Liang}, \citenamefont {Maia}, \citenamefont {Malmerberg},
  \citenamefont {Marchesini}, \citenamefont {Martin}, \citenamefont {Nass},
  \citenamefont {Neutze}, \citenamefont {Reich}, \citenamefont {Rolles},
  \citenamefont {Rudek}, \citenamefont {Rudenko}, \citenamefont {Scott},
  \citenamefont {Schlichting}, \citenamefont {Schulz}, \citenamefont {Seibert},
  \citenamefont {Shoeman}, \citenamefont {Sierra}, \citenamefont {Soltau},
  \citenamefont {Spence}, \citenamefont {Stellato}, \citenamefont {Stern},
  \citenamefont {Str{\"u}der}, \citenamefont {Ullrich}, \citenamefont {Wang},
  \citenamefont {Weidenspointner}, \citenamefont {Weierstall}, \citenamefont
  {Wunderer},\ and\ \citenamefont {Chapman}}]{Barty:NatPhoton6:35}%
  \BibitemOpen
  \bibfield  {author} {\bibinfo {author} {\bibfnamefont {Anton}\ \bibnamefont
  {Barty}}, \bibinfo {author} {\bibfnamefont {Carl}\ \bibnamefont {Caleman}},
  \bibinfo {author} {\bibfnamefont {Andrew}\ \bibnamefont {Aquila}}, \bibinfo
  {author} {\bibfnamefont {Nicusor}\ \bibnamefont {Timneanu}}, \bibinfo
  {author} {\bibfnamefont {Lukas}\ \bibnamefont {Lomb}}, \bibinfo {author}
  {\bibfnamefont {Thomas~A}\ \bibnamefont {White}}, \bibinfo {author}
  {\bibfnamefont {Jakob}\ \bibnamefont {Andreasson}}, \bibinfo {author}
  {\bibfnamefont {David}\ \bibnamefont {Arnlund}}, \bibinfo {author}
  {\bibfnamefont {Sa{\v s}a}\ \bibnamefont {Bajt}}, \bibinfo {author}
  {\bibfnamefont {Thomas R~M}\ \bibnamefont {Barends}}, \bibinfo {author}
  {\bibfnamefont {Miriam}\ \bibnamefont {Barthelmess}}, \bibinfo {author}
  {\bibfnamefont {Michael~J}\ \bibnamefont {Bogan}}, \bibinfo {author}
  {\bibfnamefont {Christoph}\ \bibnamefont {Bostedt}}, \bibinfo {author}
  {\bibfnamefont {John~D}\ \bibnamefont {Bozek}}, \bibinfo {author}
  {\bibfnamefont {Ryan}\ \bibnamefont {Coffee}}, \bibinfo {author}
  {\bibfnamefont {Nicola}\ \bibnamefont {Coppola}}, \bibinfo {author}
  {\bibfnamefont {Jan}\ \bibnamefont {Davidsson}}, \bibinfo {author}
  {\bibfnamefont {Daniel~P}\ \bibnamefont {Deponte}}, \bibinfo {author}
  {\bibfnamefont {R~Bruce}\ \bibnamefont {Doak}}, \bibinfo {author}
  {\bibfnamefont {Tomas}\ \bibnamefont {Ekeberg}}, \bibinfo {author}
  {\bibfnamefont {Veit}\ \bibnamefont {Elser}}, \bibinfo {author}
  {\bibfnamefont {Sascha~W}\ \bibnamefont {Epp}}, \bibinfo {author}
  {\bibfnamefont {Benjamin}\ \bibnamefont {Erk}}, \bibinfo {author}
  {\bibfnamefont {Holger}\ \bibnamefont {Fleckenstein}}, \bibinfo {author}
  {\bibfnamefont {Lutz}\ \bibnamefont {Foucar}}, \bibinfo {author}
  {\bibfnamefont {Petra}\ \bibnamefont {Fromme}}, \bibinfo {author}
  {\bibfnamefont {Heinz}\ \bibnamefont {Graafsma}}, \bibinfo {author}
  {\bibfnamefont {Lars}\ \bibnamefont {Gumprecht}}, \bibinfo {author}
  {\bibfnamefont {Janos}\ \bibnamefont {Hajdu}}, \bibinfo {author}
  {\bibfnamefont {Christina~Y}\ \bibnamefont {Hampton}}, \bibinfo {author}
  {\bibfnamefont {Robert}\ \bibnamefont {Hartmann}}, \bibinfo {author}
  {\bibfnamefont {Andreas}\ \bibnamefont {Hartmann}}, \bibinfo {author}
  {\bibfnamefont {G{\"u}nter}\ \bibnamefont {Hauser}}, \bibinfo {author}
  {\bibfnamefont {Helmut}\ \bibnamefont {Hirsemann}}, \bibinfo {author}
  {\bibfnamefont {Peter}\ \bibnamefont {Holl}}, \bibinfo {author}
  {\bibfnamefont {Mark~S}\ \bibnamefont {Hunter}}, \bibinfo {author}
  {\bibfnamefont {Linda}\ \bibnamefont {Johansson}}, \bibinfo {author}
  {\bibfnamefont {Stephan}\ \bibnamefont {Kassemeyer}}, \bibinfo {author}
  {\bibfnamefont {Nils}\ \bibnamefont {Kimmel}}, \bibinfo {author}
  {\bibfnamefont {Richard~A}\ \bibnamefont {Kirian}}, \bibinfo {author}
  {\bibfnamefont {Mengning}\ \bibnamefont {Liang}}, \bibinfo {author}
  {\bibfnamefont {Filipe R N~C}\ \bibnamefont {Maia}}, \bibinfo {author}
  {\bibfnamefont {Erik}\ \bibnamefont {Malmerberg}}, \bibinfo {author}
  {\bibfnamefont {Stefano}\ \bibnamefont {Marchesini}}, \bibinfo {author}
  {\bibfnamefont {Andrew~V}\ \bibnamefont {Martin}}, \bibinfo {author}
  {\bibfnamefont {Karol}\ \bibnamefont {Nass}}, \bibinfo {author}
  {\bibfnamefont {Richard}\ \bibnamefont {Neutze}}, \bibinfo {author}
  {\bibfnamefont {Christian}\ \bibnamefont {Reich}}, \bibinfo {author}
  {\bibfnamefont {Daniel}\ \bibnamefont {Rolles}}, \bibinfo {author}
  {\bibfnamefont {Benedikt}\ \bibnamefont {Rudek}}, \bibinfo {author}
  {\bibfnamefont {Artem}\ \bibnamefont {Rudenko}}, \bibinfo {author}
  {\bibfnamefont {Howard}\ \bibnamefont {Scott}}, \bibinfo {author}
  {\bibfnamefont {Ilme}\ \bibnamefont {Schlichting}}, \bibinfo {author}
  {\bibfnamefont {Joachim}\ \bibnamefont {Schulz}}, \bibinfo {author}
  {\bibfnamefont {M~Marvin}\ \bibnamefont {Seibert}}, \bibinfo {author}
  {\bibfnamefont {Robert~L}\ \bibnamefont {Shoeman}}, \bibinfo {author}
  {\bibfnamefont {Raymond~G}\ \bibnamefont {Sierra}}, \bibinfo {author}
  {\bibfnamefont {Heike}\ \bibnamefont {Soltau}}, \bibinfo {author}
  {\bibfnamefont {John C~H}\ \bibnamefont {Spence}}, \bibinfo {author}
  {\bibfnamefont {Francesco}\ \bibnamefont {Stellato}}, \bibinfo {author}
  {\bibfnamefont {Stephan}\ \bibnamefont {Stern}}, \bibinfo {author}
  {\bibfnamefont {Lothar}\ \bibnamefont {Str{\"u}der}}, \bibinfo {author}
  {\bibfnamefont {Joachim~Hermann}\ \bibnamefont {Ullrich}}, \bibinfo {author}
  {\bibfnamefont {X}~\bibnamefont {Wang}}, \bibinfo {author} {\bibfnamefont
  {Georg}\ \bibnamefont {Weidenspointner}}, \bibinfo {author} {\bibfnamefont
  {Uwe}\ \bibnamefont {Weierstall}}, \bibinfo {author} {\bibfnamefont
  {Cornelia~B}\ \bibnamefont {Wunderer}}, \ and\ \bibinfo {author}
  {\bibfnamefont {Henry~N}\ \bibnamefont {Chapman}},\ }\bibfield  {title}
  {\enquote {\bibinfo {title} {Self-terminating diffraction gates femtosecond
  x-ray nanocrystallography measurements},}\ }\href {\doibase  10.1038/nphoton.2011.297} {\bibfield  {journal} {\bibinfo  {journal} {Nat.
  Photon.}\ }\textbf {\bibinfo {volume} {6}},\ \bibinfo {pages} {35--40}
  (\bibinfo {year} {2012})}\BibitemShut {NoStop}%
\bibitem [{\citenamefont {Rudek}\ \emph {et~al.}(2012)\citenamefont {Rudek},
  \citenamefont {Son}, \citenamefont {Foucar}, \citenamefont {Epp},
  \citenamefont {Erk}, \citenamefont {Hartmann}, \citenamefont {Adolph},
  \citenamefont {Andritschke}, \citenamefont {Aquila}, \citenamefont {Berrah},
  \citenamefont {Bostedt}, \citenamefont {Bozek}, \citenamefont {Filsinger},
  \citenamefont {Gorke}, \citenamefont {Gorkhover}, \citenamefont {Graafsma},
  \citenamefont {Gumprecht}, \citenamefont {Hartmann}, \citenamefont {Hauser},
  \citenamefont {Herrmann}, \citenamefont {Hirsemann}, \citenamefont {Holl},
  \citenamefont {H{\"o}mke}, \citenamefont {Journel}, \citenamefont {Kaiser},
  \citenamefont {Kimmel}, \citenamefont {Krasniqi}, \citenamefont {K{\"u}hnel},
  \citenamefont {Matysek}, \citenamefont {Messerschmidt}, \citenamefont
  {Miesner}, \citenamefont {M{\"o}ller}, \citenamefont {Moshammer},
  \citenamefont {Nagaya}, \citenamefont {Nilsson}, \citenamefont {Potdevin},
  \citenamefont {Pietschner}, \citenamefont {Reich}, \citenamefont {Rupp},
  \citenamefont {Schaller}, \citenamefont {Schlichting}, \citenamefont
  {Schmidt}, \citenamefont {Schopper}, \citenamefont {Schorb}, \citenamefont
  {Schr{\"o}ter}, \citenamefont {Schulz}, \citenamefont {Simon}, \citenamefont
  {Soltau}, \citenamefont {Str{\"u}der}, \citenamefont {Ueda}, \citenamefont
  {Weidenspointner}, \citenamefont {Santra}, \citenamefont {Ullrich},
  \citenamefont {Rudenko},\ and\ \citenamefont
  {Rolles}}]{Rudek:NatPhoton6:858}%
  \BibitemOpen
  \bibfield  {author} {\bibinfo {author} {\bibfnamefont {Benedikt}\
  \bibnamefont {Rudek}}, \bibinfo {author} {\bibfnamefont {Sang-Kil}\
  \bibnamefont {Son}}, \bibinfo {author} {\bibfnamefont {Lutz}\ \bibnamefont
  {Foucar}}, \bibinfo {author} {\bibfnamefont {Sascha-W.}\ \bibnamefont {Epp}},
  \bibinfo {author} {\bibfnamefont {Benjamin}\ \bibnamefont {Erk}}, \bibinfo
  {author} {\bibfnamefont {Robert}\ \bibnamefont {Hartmann}}, \bibinfo {author}
  {\bibfnamefont {Marcus}\ \bibnamefont {Adolph}}, \bibinfo {author}
  {\bibfnamefont {Robert}\ \bibnamefont {Andritschke}}, \bibinfo {author}
  {\bibfnamefont {Andrew}\ \bibnamefont {Aquila}}, \bibinfo {author}
  {\bibfnamefont {Nora}\ \bibnamefont {Berrah}}, \bibinfo {author}
  {\bibfnamefont {Christoph}\ \bibnamefont {Bostedt}}, \bibinfo {author}
  {\bibfnamefont {Nicola}\ \bibnamefont {Bozek}, \bibfnamefont
  {Johnand~Coppola}}, \bibinfo {author} {\bibfnamefont {Frank}\ \bibnamefont
  {Filsinger}}, \bibinfo {author} {\bibfnamefont {Hubert}\ \bibnamefont
  {Gorke}}, \bibinfo {author} {\bibfnamefont {Tais}\ \bibnamefont {Gorkhover}},
  \bibinfo {author} {\bibfnamefont {Heinz}\ \bibnamefont {Graafsma}}, \bibinfo
  {author} {\bibfnamefont {Lars}\ \bibnamefont {Gumprecht}}, \bibinfo {author}
  {\bibfnamefont {Andreas}\ \bibnamefont {Hartmann}}, \bibinfo {author}
  {\bibfnamefont {G{\"u}nter}\ \bibnamefont {Hauser}}, \bibinfo {author}
  {\bibfnamefont {Sven}\ \bibnamefont {Herrmann}}, \bibinfo {author}
  {\bibfnamefont {Helmut}\ \bibnamefont {Hirsemann}}, \bibinfo {author}
  {\bibfnamefont {Peter}\ \bibnamefont {Holl}}, \bibinfo {author}
  {\bibfnamefont {Andre}\ \bibnamefont {H{\"o}mke}}, \bibinfo {author}
  {\bibfnamefont {Loic}\ \bibnamefont {Journel}}, \bibinfo {author}
  {\bibfnamefont {Christian}\ \bibnamefont {Kaiser}}, \bibinfo {author}
  {\bibfnamefont {Nils}\ \bibnamefont {Kimmel}}, \bibinfo {author}
  {\bibfnamefont {Faton}\ \bibnamefont {Krasniqi}}, \bibinfo {author}
  {\bibfnamefont {Kai-Uwe}\ \bibnamefont {K{\"u}hnel}}, \bibinfo {author}
  {\bibfnamefont {Michael}\ \bibnamefont {Matysek}}, \bibinfo {author}
  {\bibfnamefont {Marc}\ \bibnamefont {Messerschmidt}}, \bibinfo {author}
  {\bibfnamefont {Danilo}\ \bibnamefont {Miesner}}, \bibinfo {author}
  {\bibfnamefont {Thomas}\ \bibnamefont {M{\"o}ller}}, \bibinfo {author}
  {\bibfnamefont {Robert}\ \bibnamefont {Moshammer}}, \bibinfo {author}
  {\bibfnamefont {Kiyonobu}\ \bibnamefont {Nagaya}}, \bibinfo {author}
  {\bibfnamefont {Bjorn}\ \bibnamefont {Nilsson}}, \bibinfo {author}
  {\bibfnamefont {Guillaume}\ \bibnamefont {Potdevin}}, \bibinfo {author}
  {\bibfnamefont {Daniel}\ \bibnamefont {Pietschner}}, \bibinfo {author}
  {\bibfnamefont {Christian}\ \bibnamefont {Reich}}, \bibinfo {author}
  {\bibfnamefont {Daniela}\ \bibnamefont {Rupp}}, \bibinfo {author}
  {\bibfnamefont {Gerhard}\ \bibnamefont {Schaller}}, \bibinfo {author}
  {\bibfnamefont {Ilme}\ \bibnamefont {Schlichting}}, \bibinfo {author}
  {\bibfnamefont {Carlo}\ \bibnamefont {Schmidt}}, \bibinfo {author}
  {\bibfnamefont {Florian}\ \bibnamefont {Schopper}}, \bibinfo {author}
  {\bibfnamefont {Sebastian}\ \bibnamefont {Schorb}}, \bibinfo {author}
  {\bibfnamefont {Claus-Dieter}\ \bibnamefont {Schr{\"o}ter}}, \bibinfo
  {author} {\bibfnamefont {Joachim}\ \bibnamefont {Schulz}}, \bibinfo {author}
  {\bibfnamefont {Marc}\ \bibnamefont {Simon}}, \bibinfo {author}
  {\bibfnamefont {Heike}\ \bibnamefont {Soltau}}, \bibinfo {author}
  {\bibfnamefont {Lothar}\ \bibnamefont {Str{\"u}der}}, \bibinfo {author}
  {\bibfnamefont {Kiyoshi}\ \bibnamefont {Ueda}}, \bibinfo {author}
  {\bibfnamefont {Georg}\ \bibnamefont {Weidenspointner}}, \bibinfo {author}
  {\bibfnamefont {Robin}\ \bibnamefont {Santra}}, \bibinfo {author}
  {\bibfnamefont {Joachim}\ \bibnamefont {Ullrich}}, \bibinfo {author}
  {\bibfnamefont {Artem}\ \bibnamefont {Rudenko}}, \ and\ \bibinfo {author}
  {\bibfnamefont {Daniel}\ \bibnamefont {Rolles}},\ }\bibfield  {title}
  {\enquote {\bibinfo {title} {Ultra-efficient ionization of heavy atoms by
  intense x-ray free-electron laser pulses},}\ }\href {\doibase  10.1038/nphoton.2012.261} {\bibfield  {journal} {\bibinfo  {journal} {Nat.
  Photon.}\ }\textbf {\bibinfo {volume} {6}},\ \bibinfo {pages} {858--865}
  (\bibinfo {year} {2012})}\BibitemShut {NoStop}%
\bibitem [{\citenamefont {Barty}\ \emph {et~al.}(2013)\citenamefont {Barty},
  \citenamefont {K{\"u}pper},\ and\ \citenamefont
  {Chapman}}]{Barty:ARPC64:415}%
  \BibitemOpen
  \bibfield  {author} {\bibinfo {author} {\bibfnamefont {Anton}\ \bibnamefont
  {Barty}}, \bibinfo {author} {\bibfnamefont {Jochen}\ \bibnamefont
  {K{\"u}pper}}, \ and\ \bibinfo {author} {\bibfnamefont {Henry~N.}\
  \bibnamefont {Chapman}},\ }\bibfield  {title} {\enquote {\bibinfo {title}
  {Molecular imaging using x-ray free-electron lasers},}\ }\href {\doibase  10.1146/annurev-physchem-032511-143708} {\bibfield  {journal} {\bibinfo
  {journal} {Annu.\ Rev.\ Phys.\ Chem.}\ }\textbf {\bibinfo {volume} {64}},\
  \bibinfo {pages} {415--435} (\bibinfo {year} {2013})}\BibitemShut {NoStop}%
\bibitem [{\citenamefont {Erk}\ \emph {et~al.}(2014)\citenamefont {Erk},
  \citenamefont {Boll}, \citenamefont {Trippel}, \citenamefont {Anielski},
  \citenamefont {Foucar}, \citenamefont {Rudek}, \citenamefont {Epp},
  \citenamefont {Coffee}, \citenamefont {Carron}, \citenamefont {Schorb},
  \citenamefont {Ferguson}, \citenamefont {Swiggers}, \citenamefont {Bozek},
  \citenamefont {Simon}, \citenamefont {Marchenko}, \citenamefont {K{\"u}pper},
  \citenamefont {Schlichting}, \citenamefont {Ullrich}, \citenamefont
  {Bostedt}, \citenamefont {Rolles},\ and\ \citenamefont
  {Rudenko}}]{Erk:Science345:288}%
  \BibitemOpen
  \bibfield  {author} {\bibinfo {author} {\bibfnamefont {Benjamin}\
  \bibnamefont {Erk}}, \bibinfo {author} {\bibfnamefont {Rebecca}\ \bibnamefont
  {Boll}}, \bibinfo {author} {\bibfnamefont {Sebastian}\ \bibnamefont
  {Trippel}}, \bibinfo {author} {\bibfnamefont {Denis}\ \bibnamefont
  {Anielski}}, \bibinfo {author} {\bibfnamefont {Lutz}\ \bibnamefont {Foucar}},
  \bibinfo {author} {\bibfnamefont {Benedikt}\ \bibnamefont {Rudek}}, \bibinfo
  {author} {\bibfnamefont {Sascha~W}\ \bibnamefont {Epp}}, \bibinfo {author}
  {\bibfnamefont {Ryan}\ \bibnamefont {Coffee}}, \bibinfo {author}
  {\bibfnamefont {Sebastian}\ \bibnamefont {Carron}}, \bibinfo {author}
  {\bibfnamefont {Sebastian}\ \bibnamefont {Schorb}}, \bibinfo {author}
  {\bibfnamefont {Ken~R}\ \bibnamefont {Ferguson}}, \bibinfo {author}
  {\bibfnamefont {Michele}\ \bibnamefont {Swiggers}}, \bibinfo {author}
  {\bibfnamefont {John~D}\ \bibnamefont {Bozek}}, \bibinfo {author}
  {\bibfnamefont {Marc}\ \bibnamefont {Simon}}, \bibinfo {author}
  {\bibfnamefont {Tatiana}\ \bibnamefont {Marchenko}}, \bibinfo {author}
  {\bibfnamefont {Jochen}\ \bibnamefont {K{\"u}pper}}, \bibinfo {author}
  {\bibfnamefont {Ilme}\ \bibnamefont {Schlichting}}, \bibinfo {author}
  {\bibfnamefont {Joachim}\ \bibnamefont {Ullrich}}, \bibinfo {author}
  {\bibfnamefont {Christoph}\ \bibnamefont {Bostedt}}, \bibinfo {author}
  {\bibfnamefont {Daniel}\ \bibnamefont {Rolles}}, \ and\ \bibinfo {author}
  {\bibfnamefont {Artem}\ \bibnamefont {Rudenko}},\ }\bibfield  {title}
  {\enquote {\bibinfo {title} {Imaging charge transfer in iodomethane upon
  x-ray photoabsorption.}}\ }\href {\doibase 10.1126/science.1253607}
  {\bibfield  {journal} {\bibinfo  {journal} {Science}\ }\textbf {\bibinfo
  {volume} {345}},\ \bibinfo {pages} {288--291} (\bibinfo {year}
  {2014})}\BibitemShut {NoStop}%
\bibitem [{\citenamefont {Huang}\ and\ \citenamefont
  {Kim}(2007)}]{Huang:PRSTAB10:034801}%
  \BibitemOpen
  \bibfield  {author} {\bibinfo {author} {\bibfnamefont {Zhirong}\ \bibnamefont
  {Huang}}\ and\ \bibinfo {author} {\bibfnamefont {Kwang~Je}\ \bibnamefont
  {Kim}},\ }\bibfield  {title} {\enquote {\bibinfo {title} {{Review of x-ray
  free-electron laser theory}},}\ }\href {\doibase  10.1103/PhysRevSTAB.10.034801} {\bibfield  {journal} {\bibinfo  {journal}
  {Phys.\ Rev.\ ST\ Accel.\ Beams}\ }\textbf {\bibinfo {volume} {10}},\
  \bibinfo {pages} {1--26} (\bibinfo {year} {2007})}\BibitemShut {NoStop}%
\bibitem [{\citenamefont {Ihee}\ \emph {et~al.}(2001)\citenamefont {Ihee},
  \citenamefont {Lobastov}, \citenamefont {Gomez}, \citenamefont {Goodson},
  \citenamefont {Srinivasan}, \citenamefont {Ruan},\ and\ \citenamefont
  {Zewail}}]{Ihee:Science291:458}%
  \BibitemOpen
  \bibfield  {author} {\bibinfo {author} {\bibfnamefont {H}~\bibnamefont
  {Ihee}}, \bibinfo {author} {\bibfnamefont {VA}~\bibnamefont {Lobastov}},
  \bibinfo {author} {\bibfnamefont {UM}~\bibnamefont {Gomez}}, \bibinfo
  {author} {\bibfnamefont {BM}~\bibnamefont {Goodson}}, \bibinfo {author}
  {\bibfnamefont {R}~\bibnamefont {Srinivasan}}, \bibinfo {author}
  {\bibfnamefont {CY}~\bibnamefont {Ruan}}, \ and\ \bibinfo {author}
  {\bibfnamefont {Ahmed~H}\ \bibnamefont {Zewail}},\ }\bibfield  {title}
  {\enquote {\bibinfo {title} {Direct imaging of transient molecular structures
  with ultrafast diffraction},}\ }\href {\doibase 10.1126/science.291.5503.458}
  {\bibfield  {journal} {\bibinfo  {journal} {Science}\ }\textbf {\bibinfo
  {volume} {291}},\ \bibinfo {pages} {458--462} (\bibinfo {year}
  {2001})}\BibitemShut {NoStop}%
\bibitem [{\citenamefont {Siwick}\ \emph {et~al.}(2003)\citenamefont {Siwick},
  \citenamefont {Dwyer}, \citenamefont {Jordan},\ and\ \citenamefont
  {Miller}}]{Siwick:Science302:1382}%
  \BibitemOpen
  \bibfield  {author} {\bibinfo {author} {\bibfnamefont {B~J}\ \bibnamefont
  {Siwick}}, \bibinfo {author} {\bibfnamefont {J~R}\ \bibnamefont {Dwyer}},
  \bibinfo {author} {\bibfnamefont {R~E}\ \bibnamefont {Jordan}}, \ and\
  \bibinfo {author} {\bibfnamefont {R~J~Dwayne}\ \bibnamefont {Miller}},\
  }\bibfield  {title} {\enquote {\bibinfo {title} {An atomic-level view of
  melting using femtosecond electron diffraction},}\ }\href {\doibase  10.1126/science.1090052} {\bibfield  {journal} {\bibinfo  {journal}
  {Science}\ }\textbf {\bibinfo {volume} {302}},\ \bibinfo {pages} {1382--1385}
  (\bibinfo {year} {2003})}\BibitemShut {NoStop}%
\bibitem [{\citenamefont {Gulde}\ \emph {et~al.}(2014)\citenamefont {Gulde},
  \citenamefont {Schweda}, \citenamefont {Storeck}, \citenamefont {Maiti},
  \citenamefont {Yu}, \citenamefont {Wodtke}, \citenamefont {Sch{\"a}fer},\
  and\ \citenamefont {Ropers}}]{Gulde:Science345:200}%
  \BibitemOpen
  \bibfield  {author} {\bibinfo {author} {\bibfnamefont {Max}\ \bibnamefont
  {Gulde}}, \bibinfo {author} {\bibfnamefont {Simon}\ \bibnamefont {Schweda}},
  \bibinfo {author} {\bibfnamefont {Gero}\ \bibnamefont {Storeck}}, \bibinfo
  {author} {\bibfnamefont {Manisankar}\ \bibnamefont {Maiti}}, \bibinfo
  {author} {\bibfnamefont {Hak~Ki}\ \bibnamefont {Yu}}, \bibinfo {author}
  {\bibfnamefont {Alec~M}\ \bibnamefont {Wodtke}}, \bibinfo {author}
  {\bibfnamefont {Sascha}\ \bibnamefont {Sch{\"a}fer}}, \ and\ \bibinfo
  {author} {\bibfnamefont {Claus}\ \bibnamefont {Ropers}},\ }\bibfield  {title}
  {\enquote {\bibinfo {title} {Ultrafast low-energy electron diffraction in
  transmission resolves polymer/graphene superstructure dynamics},}\ }\href
  {\doibase 10.1126/science.1250658} {\bibfield  {journal} {\bibinfo  {journal}
  {Science}\ }\textbf {\bibinfo {volume} {345}},\ \bibinfo {pages} {200--204}
  (\bibinfo {year} {2014})}\BibitemShut {NoStop}%
\bibitem [{\citenamefont {Yang}\ \emph {et~al.}(2016)\citenamefont {Yang},
  \citenamefont {Guehr}, \citenamefont {Shen}, \citenamefont {Li},
  \citenamefont {Vecchione}, \citenamefont {Coffee}, \citenamefont {Corbett},
  \citenamefont {Fry}, \citenamefont {Hartmann}, \citenamefont {Hast},
  \citenamefont {Hegazy}, \citenamefont {Jobe}, \citenamefont {Makasyuk},
  \citenamefont {Robinson}, \citenamefont {Robinson}, \citenamefont {Vetter},
  \citenamefont {Weathersby}, \citenamefont {Yoneda}, \citenamefont {Wang},\
  and\ \citenamefont {Centurion}}]{Yang:PRL117:153002}%
  \BibitemOpen
  \bibfield  {author} {\bibinfo {author} {\bibfnamefont {Jie}\ \bibnamefont
  {Yang}}, \bibinfo {author} {\bibfnamefont {Markus}\ \bibnamefont {Guehr}},
  \bibinfo {author} {\bibfnamefont {Xiaozhe}\ \bibnamefont {Shen}}, \bibinfo
  {author} {\bibfnamefont {Renkai}\ \bibnamefont {Li}}, \bibinfo {author}
  {\bibfnamefont {Theodore}\ \bibnamefont {Vecchione}}, \bibinfo {author}
  {\bibfnamefont {Ryan}\ \bibnamefont {Coffee}}, \bibinfo {author}
  {\bibfnamefont {Jeff}\ \bibnamefont {Corbett}}, \bibinfo {author}
  {\bibfnamefont {Alan}\ \bibnamefont {Fry}}, \bibinfo {author} {\bibfnamefont
  {Nick}\ \bibnamefont {Hartmann}}, \bibinfo {author} {\bibfnamefont {Carsten}\
  \bibnamefont {Hast}}, \bibinfo {author} {\bibfnamefont {Kareem}\ \bibnamefont
  {Hegazy}}, \bibinfo {author} {\bibfnamefont {Keith}\ \bibnamefont {Jobe}},
  \bibinfo {author} {\bibfnamefont {Igor}\ \bibnamefont {Makasyuk}}, \bibinfo
  {author} {\bibfnamefont {Joseph}\ \bibnamefont {Robinson}}, \bibinfo {author}
  {\bibfnamefont {Matthew~S}\ \bibnamefont {Robinson}}, \bibinfo {author}
  {\bibfnamefont {Sharon}\ \bibnamefont {Vetter}}, \bibinfo {author}
  {\bibfnamefont {Stephen}\ \bibnamefont {Weathersby}}, \bibinfo {author}
  {\bibfnamefont {Charles}\ \bibnamefont {Yoneda}}, \bibinfo {author}
  {\bibfnamefont {Xijie}\ \bibnamefont {Wang}}, \ and\ \bibinfo {author}
  {\bibfnamefont {Martin}\ \bibnamefont {Centurion}},\ }\bibfield  {title}
  {\enquote {\bibinfo {title} {Diffractive imaging of coherent nuclear motion
  in isolated molecules},}\ }\href {\doibase 10.1103/PhysRevLett.117.153002}
  {\bibfield  {journal} {\bibinfo  {journal} {Phys.\ Rev.\ Lett.}\ }\textbf
  {\bibinfo {volume} {117}},\ \bibinfo {pages} {153002} (\bibinfo {year}
  {2016})}\BibitemShut {NoStop}%
\bibitem [{\citenamefont {Kr{\"u}ger}\ \emph {et~al.}(2011)\citenamefont
  {Kr{\"u}ger}, \citenamefont {Schenk},\ and\ \citenamefont
  {Hommelhoff}}]{Kruger:Nature475:78}%
  \BibitemOpen
  \bibfield  {author} {\bibinfo {author} {\bibfnamefont {Michael}\ \bibnamefont
  {Kr{\"u}ger}}, \bibinfo {author} {\bibfnamefont {Markus}\ \bibnamefont
  {Schenk}}, \ and\ \bibinfo {author} {\bibfnamefont {Peter}\ \bibnamefont
  {Hommelhoff}},\ }\bibfield  {title} {\enquote {\bibinfo {title} {Attosecond
  control of electrons emitted from a nanoscale metal tip.}}\ }\href {\doibase  10.1038/nature10196} {\bibfield  {journal} {\bibinfo  {journal} {Nature}\
  }\textbf {\bibinfo {volume} {475}},\ \bibinfo {pages} {78--81} (\bibinfo
  {year} {2011})}\BibitemShut {NoStop}%
\bibitem [{\citenamefont {Herink}\ \emph {et~al.}(2012)\citenamefont {Herink},
  \citenamefont {Solli}, \citenamefont {Gulde},\ and\ \citenamefont
  {Ropers}}]{Herink:Nature483:193}%
  \BibitemOpen
  \bibfield  {author} {\bibinfo {author} {\bibfnamefont {G}~\bibnamefont
  {Herink}}, \bibinfo {author} {\bibfnamefont {D~R}\ \bibnamefont {Solli}},
  \bibinfo {author} {\bibfnamefont {M}~\bibnamefont {Gulde}}, \ and\ \bibinfo
  {author} {\bibfnamefont {C}~\bibnamefont {Ropers}},\ }\bibfield  {title}
  {\enquote {\bibinfo {title} {Field-driven photoemission from nanostructures
  quenches the quiver motion.}}\ }\href {\doibase 10.1038/nature10878}
  {\bibfield  {journal} {\bibinfo  {journal} {Nature}\ }\textbf {\bibinfo
  {volume} {483}},\ \bibinfo {pages} {190--193} (\bibinfo {year}
  {2012})}\BibitemShut {NoStop}%
\bibitem [{\citenamefont {Mustonen}\ \emph {et~al.}(2011)\citenamefont
  {Mustonen}, \citenamefont {Beaud}, \citenamefont {Kirk}, \citenamefont
  {Feurer},\ and\ \citenamefont {Tsujino}}]{Mustonen:APL99:103504}%
  \BibitemOpen
  \bibfield  {author} {\bibinfo {author} {\bibfnamefont {Anna}\ \bibnamefont
  {Mustonen}}, \bibinfo {author} {\bibfnamefont {Paul}\ \bibnamefont {Beaud}},
  \bibinfo {author} {\bibfnamefont {Eugenie}\ \bibnamefont {Kirk}}, \bibinfo
  {author} {\bibfnamefont {Thomas}\ \bibnamefont {Feurer}}, \ and\ \bibinfo
  {author} {\bibfnamefont {Soichiro}\ \bibnamefont {Tsujino}},\ }\bibfield
  {title} {\enquote {\bibinfo {title} {Five picocoulomb electron bunch
  generation by ultrafast laser-induced field emission from metallic nano-tip
  arrays},}\ }\href {\doibase 10.1063/1.3631634} {\bibfield  {journal}
  {\bibinfo  {journal} {Astrophys.\ Lett.\ \& Comm.}\ }\textbf {\bibinfo
  {volume} {99}},\ \bibinfo {pages} {103504} (\bibinfo {year}
  {2011})}\BibitemShut {NoStop}%
\bibitem [{\citenamefont {Putnam}\ \emph {et~al.}(2016)\citenamefont {Putnam},
  \citenamefont {Hobbs}, \citenamefont {Keathley}, \citenamefont {Berggren},\
  and\ \citenamefont {K{\"a}rtner}}]{Putnam:NatPhys13:335}%
  \BibitemOpen
  \bibfield  {author} {\bibinfo {author} {\bibfnamefont {William~P}\
  \bibnamefont {Putnam}}, \bibinfo {author} {\bibfnamefont {Richard~G}\
  \bibnamefont {Hobbs}}, \bibinfo {author} {\bibfnamefont {Phillip~D}\
  \bibnamefont {Keathley}}, \bibinfo {author} {\bibfnamefont {Karl~K}\
  \bibnamefont {Berggren}}, \ and\ \bibinfo {author} {\bibfnamefont {Franz~X}\
  \bibnamefont {K{\"a}rtner}},\ }\bibfield  {title} {\enquote {\bibinfo {title}
  {Optical-field-controlled photoemission from plasmonic nanoparticles},}\
  }\href {\doibase 10.1038/nphys3978} {\bibfield  {journal} {\bibinfo
  {journal} {Nat. Phys.}\ }\textbf {\bibinfo {volume} {13}},\ \bibinfo {pages}
  {335--339} (\bibinfo {year} {2016})}\BibitemShut {NoStop}%
\bibitem [{\citenamefont {Tsujino}\ \emph {et~al.}(2016)\citenamefont
  {Tsujino}, \citenamefont {{Das Kanungo}}, \citenamefont {Monshipouri},
  \citenamefont {Lee},\ and\ \citenamefont {Miller}}]{Tsujino:NatComm7:13976}%
  \BibitemOpen
  \bibfield  {author} {\bibinfo {author} {\bibfnamefont {Soichiro}\
  \bibnamefont {Tsujino}}, \bibinfo {author} {\bibfnamefont {Prat}\
  \bibnamefont {{Das Kanungo}}}, \bibinfo {author} {\bibfnamefont {Mahta}\
  \bibnamefont {Monshipouri}}, \bibinfo {author} {\bibfnamefont {Chiwon}\
  \bibnamefont {Lee}}, \ and\ \bibinfo {author} {\bibfnamefont {R.J.~Dwayne}\
  \bibnamefont {Miller}},\ }\bibfield  {title} {\enquote {\bibinfo {title}
  {Measurement of transverse emittance and coherence of double-gate field
  emitter array cathodes},}\ }\href {\doibase 10.1038/ncomms13976} {\bibfield
  {journal} {\bibinfo  {journal} {Nat. Commun.}\ }\textbf {\bibinfo {volume}
  {7}},\ \bibinfo {pages} {13976} (\bibinfo {year} {2016})}\BibitemShut
  {NoStop}%
\bibitem [{\citenamefont {K{\"{a}}rtner}\ \emph {et~al.}(2016)\citenamefont
  {K{\"{a}}rtner}, \citenamefont {Ahr}, \citenamefont {Calendron},
  \citenamefont {{\c{C}}ankaya}, \citenamefont {Carbajo}, \citenamefont
  {Chang}, \citenamefont {Cirmi}, \citenamefont {D{\"{o}}rner}, \citenamefont
  {Dorda}, \citenamefont {Fallahi}, \citenamefont {Hartin}, \citenamefont
  {Hemmer}, \citenamefont {Hobbs}, \citenamefont {Hua}, \citenamefont {Huang},
  \citenamefont {Letrun}, \citenamefont {Matlis}, \citenamefont {Mazalova},
  \citenamefont {M{\"{u}}cke}, \citenamefont {Nanni}, \citenamefont {Putnam},
  \citenamefont {Ravi}, \citenamefont {Reichert}, \citenamefont {Sarrou},
  \citenamefont {Wu}, \citenamefont {Yahaghi}, \citenamefont {Ye},
  \citenamefont {Zapata}, \citenamefont {Zhang}, \citenamefont {Zhou},
  \citenamefont {Miller}, \citenamefont {Berggren}, \citenamefont {Graafsma},
  \citenamefont {Meents}, \citenamefont {Assmann}, \citenamefont {Chapman},\
  and\ \citenamefont {Fromme}}]{Kaertner:NIMA829:24}%
  \BibitemOpen
  \bibfield  {author} {\bibinfo {author} {\bibfnamefont {F.~X.}\ \bibnamefont
  {K{\"{a}}rtner}}, \bibinfo {author} {\bibfnamefont {F.}~\bibnamefont {Ahr}},
  \bibinfo {author} {\bibfnamefont {A.~L.}\ \bibnamefont {Calendron}}, \bibinfo
  {author} {\bibfnamefont {H.}~\bibnamefont {{\c{C}}ankaya}}, \bibinfo {author}
  {\bibfnamefont {S.}~\bibnamefont {Carbajo}}, \bibinfo {author} {\bibfnamefont
  {G.}~\bibnamefont {Chang}}, \bibinfo {author} {\bibfnamefont
  {G.}~\bibnamefont {Cirmi}}, \bibinfo {author} {\bibfnamefont
  {K.}~\bibnamefont {D{\"{o}}rner}}, \bibinfo {author} {\bibfnamefont
  {U.}~\bibnamefont {Dorda}}, \bibinfo {author} {\bibfnamefont
  {A.}~\bibnamefont {Fallahi}}, \bibinfo {author} {\bibfnamefont
  {A.}~\bibnamefont {Hartin}}, \bibinfo {author} {\bibfnamefont
  {M.}~\bibnamefont {Hemmer}}, \bibinfo {author} {\bibfnamefont
  {R.}~\bibnamefont {Hobbs}}, \bibinfo {author} {\bibfnamefont
  {Y.}~\bibnamefont {Hua}}, \bibinfo {author} {\bibfnamefont {W.~R.}\
  \bibnamefont {Huang}}, \bibinfo {author} {\bibfnamefont {R.}~\bibnamefont
  {Letrun}}, \bibinfo {author} {\bibfnamefont {N.}~\bibnamefont {Matlis}},
  \bibinfo {author} {\bibfnamefont {V.}~\bibnamefont {Mazalova}}, \bibinfo
  {author} {\bibfnamefont {O.~D.}\ \bibnamefont {M{\"{u}}cke}}, \bibinfo
  {author} {\bibfnamefont {E.}~\bibnamefont {Nanni}}, \bibinfo {author}
  {\bibfnamefont {W.}~\bibnamefont {Putnam}}, \bibinfo {author} {\bibfnamefont
  {K.}~\bibnamefont {Ravi}}, \bibinfo {author} {\bibfnamefont {F.}~\bibnamefont
  {Reichert}}, \bibinfo {author} {\bibfnamefont {I.}~\bibnamefont {Sarrou}},
  \bibinfo {author} {\bibfnamefont {X.}~\bibnamefont {Wu}}, \bibinfo {author}
  {\bibfnamefont {A.}~\bibnamefont {Yahaghi}}, \bibinfo {author} {\bibfnamefont
  {H.}~\bibnamefont {Ye}}, \bibinfo {author} {\bibfnamefont {L.}~\bibnamefont
  {Zapata}}, \bibinfo {author} {\bibfnamefont {D.}~\bibnamefont {Zhang}},
  \bibinfo {author} {\bibfnamefont {C.}~\bibnamefont {Zhou}}, \bibinfo {author}
  {\bibfnamefont {R.~J~D}\ \bibnamefont {Miller}}, \bibinfo {author}
  {\bibfnamefont {K.~K.}\ \bibnamefont {Berggren}}, \bibinfo {author}
  {\bibfnamefont {H.}~\bibnamefont {Graafsma}}, \bibinfo {author}
  {\bibfnamefont {A.}~\bibnamefont {Meents}}, \bibinfo {author} {\bibfnamefont
  {R.~W.}\ \bibnamefont {Assmann}}, \bibinfo {author} {\bibfnamefont {H.~N.}\
  \bibnamefont {Chapman}}, \ and\ \bibinfo {author} {\bibfnamefont
  {P.}~\bibnamefont {Fromme}},\ }\bibfield  {title} {\enquote {\bibinfo {title}
  {Axsis: Exploring the frontiers in attosecond x-ray science, imaging and
  spectroscopy},}\ }\href {\doibase 10.1016/j.nima.2016.02.080} {\bibfield
  {journal} {\bibinfo  {journal} {Nucl.\ Instrum.\ Meth.\ A}\ }\textbf
  {\bibinfo {volume} {829}},\ \bibinfo {pages} {24--29} (\bibinfo {year}
  {2016})}\BibitemShut {NoStop}%
\bibitem [{\citenamefont {Berglund}\ and\ \citenamefont
  {Spicer}(1964{\natexlab{a}})}]{Berglund:PR136:A1030}%
  \BibitemOpen
  \bibfield  {author} {\bibinfo {author} {\bibfnamefont {C.~N.}\ \bibnamefont
  {Berglund}}\ and\ \bibinfo {author} {\bibfnamefont {W.~E.}\ \bibnamefont
  {Spicer}},\ }\bibfield  {title} {\enquote {\bibinfo {title} {Photoemission
  studies of copper and silver: Theory},}\ }\href
  {http://dx.doi.org/10.1103/PhysRev.136.A1030} {\bibfield  {journal} {\bibinfo
   {journal} {Phys.\ Rev.}\ }\textbf {\bibinfo {volume} {136}},\ \bibinfo
  {pages} {A1030} (\bibinfo {year} {1964}{\natexlab{a}})}\BibitemShut {NoStop}%
\bibitem [{\citenamefont {Berglund}\ and\ \citenamefont
  {Spicer}(1964{\natexlab{b}})}]{Berglund:PR136:A1044}%
  \BibitemOpen
  \bibfield  {author} {\bibinfo {author} {\bibfnamefont {C.~N.}\ \bibnamefont
  {Berglund}}\ and\ \bibinfo {author} {\bibfnamefont {W.~E.}\ \bibnamefont
  {Spicer}},\ }\bibfield  {title} {\enquote {\bibinfo {title} {Photoemission
  studies of copper and silver: Experiment},}\ }\href {\doibase  10.1103/PhysRev.136.A1044} {\bibfield  {journal} {\bibinfo  {journal} {Phys.\
  Rev.}\ }\textbf {\bibinfo {volume} {136}},\ \bibinfo {pages} {A1044--A1064}
  (\bibinfo {year} {1964}{\natexlab{b}})}\BibitemShut {NoStop}%
\bibitem [{\citenamefont {Krolikowski}\ and\ \citenamefont
  {Spicer}(1969)}]{Krolikowski:PRB185:882}%
  \BibitemOpen
  \bibfield  {author} {\bibinfo {author} {\bibfnamefont {W.~F.}\ \bibnamefont
  {Krolikowski}}\ and\ \bibinfo {author} {\bibfnamefont {W.~E.}\ \bibnamefont
  {Spicer}},\ }\bibfield  {title} {\enquote {\bibinfo {title} {Photoemission
  studies of the noble metals. i. copper},}\ }\href {\doibase  10.1103/PhysRev.185.882} {\bibfield  {journal} {\bibinfo  {journal} {Phys.\
  Rev.\ B}\ }\textbf {\bibinfo {volume} {185}},\ \bibinfo {pages} {882--900}
  (\bibinfo {year} {1969})}\BibitemShut {NoStop}%
\bibitem [{\citenamefont {Krolikowski}\ and\ \citenamefont
  {Spicer}(1970)}]{Krolikowski:PRB1:478}%
  \BibitemOpen
  \bibfield  {author} {\bibinfo {author} {\bibfnamefont {W.~F.}\ \bibnamefont
  {Krolikowski}}\ and\ \bibinfo {author} {\bibfnamefont {W.~E.}\ \bibnamefont
  {Spicer}},\ }\bibfield  {title} {\enquote {\bibinfo {title} {Photoemission
  studies of the noble metals. ii. gold},}\ }\href {\doibase  10.1103/PhysRevB.1.478} {\bibfield  {journal} {\bibinfo  {journal} {Phys.\
  Rev.\ B}\ }\textbf {\bibinfo {volume} {1}},\ \bibinfo {pages} {478--487}
  (\bibinfo {year} {1970})}\BibitemShut {NoStop}%
\bibitem [{\citenamefont {Eppink}\ and\ \citenamefont
  {Parker}(1997)}]{Eppink:RSI68:3477}%
  \BibitemOpen
  \bibfield  {author} {\bibinfo {author} {\bibfnamefont {Andr\'{e} T. J.~B.}\
  \bibnamefont {Eppink}}\ and\ \bibinfo {author} {\bibfnamefont {David~H.}\
  \bibnamefont {Parker}},\ }\bibfield  {title} {\enquote {\bibinfo {title}
  {Velocity map imaging of ions and electrons using electrostatic lenses:
  Application in photoelectron and photofragment ion imaging of molecular
  oxygen},}\ }\href {\doibase 10.1063/1.1148310} {\bibfield  {journal}
  {\bibinfo  {journal} {Rev.\ Sci.\ Instrum.}\ }\textbf {\bibinfo {volume}
  {68}},\ \bibinfo {pages} {3477--3484} (\bibinfo {year} {1997})}\BibitemShut
  {NoStop}%
\bibitem [{\citenamefont {Chichinin}\ \emph {et~al.}(2009)\citenamefont
  {Chichinin}, \citenamefont {Gericke}, \citenamefont {Kauczok},\ and\
  \citenamefont {Maul}}]{Chichinin:IRPC28:607}%
  \BibitemOpen
  \bibfield  {author} {\bibinfo {author} {\bibfnamefont {A~I}\ \bibnamefont
  {Chichinin}}, \bibinfo {author} {\bibfnamefont {K~H}\ \bibnamefont
  {Gericke}}, \bibinfo {author} {\bibfnamefont {S}~\bibnamefont {Kauczok}}, \
  and\ \bibinfo {author} {\bibfnamefont {C}~\bibnamefont {Maul}},\ }\bibfield
  {title} {\enquote {\bibinfo {title} {Imaging chemical reactions --- 3d
  velocity mapping},}\ }\href {\doibase 10.1080/01442350903235045} {\bibfield
  {journal} {\bibinfo  {journal} {Int.\ Rev.\ Phys.\ Chem.}\ }\textbf {\bibinfo
  {volume} {28}},\ \bibinfo {pages} {607--680} (\bibinfo {year}
  {2009})}\BibitemShut {NoStop}%
\bibitem [{\citenamefont {Stei}\ \emph {et~al.}(2013)\citenamefont {Stei},
  \citenamefont {von Vangerow}, \citenamefont {Otto}, \citenamefont {Kelkar},
  \citenamefont {Carrascosa}, \citenamefont {Best},\ and\ \citenamefont
  {Wester}}]{Stei:JCP138:214201}%
  \BibitemOpen
  \bibfield  {author} {\bibinfo {author} {\bibfnamefont {M.}~\bibnamefont
  {Stei}}, \bibinfo {author} {\bibfnamefont {J.}~\bibnamefont {von Vangerow}},
  \bibinfo {author} {\bibfnamefont {R.}~\bibnamefont {Otto}}, \bibinfo {author}
  {\bibfnamefont {A.~H.}\ \bibnamefont {Kelkar}}, \bibinfo {author}
  {\bibfnamefont {E.}~\bibnamefont {Carrascosa}}, \bibinfo {author}
  {\bibfnamefont {T.}~\bibnamefont {Best}}, \ and\ \bibinfo {author}
  {\bibfnamefont {R.}~\bibnamefont {Wester}},\ }\bibfield  {title} {\enquote
  {\bibinfo {title} {High resolution spatial map imaging of a gaseous
  target},}\ }\href {\doibase 10.1063/1.4807482} {\bibfield  {journal}
  {\bibinfo  {journal} {J.\ Chem.\ Phys.}\ }\textbf {\bibinfo {volume} {138}},\
  \bibinfo {pages} {214201} (\bibinfo {year} {2013})}\BibitemShut {NoStop}%
\bibitem [{\citenamefont {Petite}\ \emph {et~al.}(1992)\citenamefont {Petite},
  \citenamefont {Agostini}, \citenamefont {Trainham}, \citenamefont {Mevel},\
  and\ \citenamefont {Martin}}]{Petite:PRB45:12210}%
  \BibitemOpen
  \bibfield  {author} {\bibinfo {author} {\bibfnamefont {Guillaume}\
  \bibnamefont {Petite}}, \bibinfo {author} {\bibfnamefont {Pierre}\
  \bibnamefont {Agostini}}, \bibinfo {author} {\bibfnamefont {Rusty}\
  \bibnamefont {Trainham}}, \bibinfo {author} {\bibfnamefont {Eric}\
  \bibnamefont {Mevel}}, \ and\ \bibinfo {author} {\bibfnamefont {Philippe}\
  \bibnamefont {Martin}},\ }\bibfield  {title} {\enquote {\bibinfo {title}
  {Electron emission from metals under laser irradiation},}\ }\href {\doibase  10.1103/PhysRevB.45.12210} {\bibfield  {journal} {\bibinfo  {journal} {Phys.\
  Rev.\ B}\ }\textbf {\bibinfo {volume} {45}},\ \bibinfo {pages} {12 210 -- 12
  217} (\bibinfo {year} {1992})}\BibitemShut {NoStop}%
\bibitem [{\citenamefont {M{\"u}ller}\ \emph {et~al.}(2015)\citenamefont
  {M{\"u}ller}, \citenamefont {Trippel}, \citenamefont {D{\l}ugo{\l}\k{e}cki},\
  and\ \citenamefont {K{\"u}pper}}]{Mueller:JPB48:244001}%
  \BibitemOpen
  \bibfield  {author} {\bibinfo {author} {\bibfnamefont {Nele L.~M.}\
  \bibnamefont {M{\"u}ller}}, \bibinfo {author} {\bibfnamefont {Sebastian}\
  \bibnamefont {Trippel}}, \bibinfo {author} {\bibfnamefont {Karol}\
  \bibnamefont {D{\l}ugo{\l}\k{e}cki}}, \ and\ \bibinfo {author} {\bibfnamefont
  {Jochen}\ \bibnamefont {K{\"u}pper}},\ }\bibfield  {title} {\enquote
  {\bibinfo {title} {Electron gun for diffraction experiments on controlled
  molecules},}\ }\href {\doibase 10.1088/0953-4075/48/24/244001} {\bibfield
  {journal} {\bibinfo  {journal} {J.\ Phys.\ B}\ }\textbf {\bibinfo {volume}
  {48}},\ \bibinfo {pages} {244001} (\bibinfo {year} {2015})},\ \Eprint
  {http://arxiv.org/abs/1507.02530} {arXiv:1507.02530 [physics]} \BibitemShut
  {NoStop}%
\bibitem [{\citenamefont {{Scientific Instrument Services Inc.,
  USA}}(2011)}]{Simion:8.1}%
  \BibitemOpen
  \bibfield  {author} {\bibinfo {author} {\bibnamefont {{Scientific Instrument
  Services Inc., USA}}},\ }\href@noop {} {\enquote {\bibinfo {title} {Simion
  8.1},}\ } (\bibinfo {year} {2011}),\ \bibinfo {note}
  {{URL}:~\url{http://simion.com}}\BibitemShut {NoStop}%
\bibitem [{\citenamefont {Bechtel}\ \emph {et~al.}(1975)\citenamefont
  {Bechtel}, \citenamefont {Smith},\ and\ \citenamefont
  {Bloembergen}}]{Bechtel:OPTCOMM13:56}%
  \BibitemOpen
  \bibfield  {author} {\bibinfo {author} {\bibfnamefont {J.~H.}\ \bibnamefont
  {Bechtel}}, \bibinfo {author} {\bibfnamefont {W.~L.}\ \bibnamefont {Smith}},
  \ and\ \bibinfo {author} {\bibfnamefont {N.}~\bibnamefont {Bloembergen}},\
  }\bibfield  {title} {\enquote {\bibinfo {title} {{Four-photon photoemission
  from tungsten}},}\ }\href {\doibase 10.1016/0030-4018(75)90156-X} {\bibfield
  {journal} {\bibinfo  {journal} {Opt.\ Comm.}\ }\textbf {\bibinfo {volume}
  {13}},\ \bibinfo {pages} {56--59} (\bibinfo {year} {1975})}\BibitemShut
  {NoStop}%
\bibitem [{\citenamefont {Damascelli}\ \emph {et~al.}(1996)\citenamefont
  {Damascelli}, \citenamefont {Gabetta}, \citenamefont {Lumachi}, \citenamefont
  {Fini},\ and\ \citenamefont {Parmigiani}}]{Damascelli:PRB54:6031}%
  \BibitemOpen
  \bibfield  {author} {\bibinfo {author} {\bibfnamefont {A}~\bibnamefont
  {Damascelli}}, \bibinfo {author} {\bibfnamefont {G}~\bibnamefont {Gabetta}},
  \bibinfo {author} {\bibfnamefont {A}~\bibnamefont {Lumachi}}, \bibinfo
  {author} {\bibfnamefont {L}~\bibnamefont {Fini}}, \ and\ \bibinfo {author}
  {\bibfnamefont {F}~\bibnamefont {Parmigiani}},\ }\bibfield  {title} {\enquote
  {\bibinfo {title} {{Multiphoton electron emission from Cu and W: An
  angle-resolved study.}}}\ }\href {\doibase 10.1103/PhysRevB.54.6031}
  {\bibfield  {journal} {\bibinfo  {journal} {Phys.\ Rev.\ B}\ }\textbf
  {\bibinfo {volume} {54}},\ \bibinfo {pages} {6031--6034} (\bibinfo {year}
  {1996})}\BibitemShut {NoStop}%
\bibitem [{\citenamefont {Lide}(2003)}]{Lide:CRCHandbookChemPhys:2003}%
  \BibitemOpen
  \bibfield  {author} {\bibinfo {author} {\bibfnamefont {David~R.}\
  \bibnamefont {Lide}},\ }\href {http://hbcponline.com} {\emph {\bibinfo
  {title} {{CRC} Handbook of Chemistry and Physics}}},\ \bibinfo {edition}
  {84th}\ ed.\ (\bibinfo  {publisher} {CRC Press},\ \bibinfo {year}
  {2003})\BibitemShut {NoStop}%
\bibitem [{\citenamefont {Polyanskiy}()}]{Polyanskiy:RID:2017}%
  \BibitemOpen
  \bibfield  {author} {\bibinfo {author} {\bibfnamefont {Mikhail~N.}\
  \bibnamefont {Polyanskiy}},\ }\href@noop {} {\enquote {\bibinfo {title}
  {Refractive index database},}\ }\bibinfo {howpublished}
  {\url{https://refractiveindex.info}},\ \bibinfo {note} {accessed on
  24.~July~2017}\BibitemShut {NoStop}%
\bibitem [{\citenamefont {Dasch}(1992)}]{Dasch:AO31:1146}%
  \BibitemOpen
  \bibfield  {author} {\bibinfo {author} {\bibfnamefont {Cameron~J.}\
  \bibnamefont {Dasch}},\ }\bibfield  {title} {\enquote {\bibinfo {title}
  {{One-dimensional tomography: a comparison of Abel, onion-peeling, and
  filtered backprojection methods}},}\ }\href {\doibase 10.1364/AO.31.001146}
  {\bibfield  {journal} {\bibinfo  {journal} {Applied Optics}\ }\textbf
  {\bibinfo {volume} {31}},\ \bibinfo {pages} {1146} (\bibinfo {year}
  {1992})}\BibitemShut {NoStop}%
\bibitem [{\citenamefont {Dowell}\ and\ \citenamefont
  {Schmerge}(2009)}]{Dowell:PRSTAB12:074201}%
  \BibitemOpen
  \bibfield  {author} {\bibinfo {author} {\bibfnamefont {David~H.}\
  \bibnamefont {Dowell}}\ and\ \bibinfo {author} {\bibfnamefont {John~F.}\
  \bibnamefont {Schmerge}},\ }\bibfield  {title} {\enquote {\bibinfo {title}
  {Quantum efficiency and thermal emittance of metal photocathodes},}\ }\href
  {\doibase 10.1103/PhysRevSTAB.12.074201} {\bibfield  {journal} {\bibinfo
  {journal} {Phys.\ Rev.\ ST\ Accel.\ Beams}\ }\textbf {\bibinfo {volume}
  {12}},\ \bibinfo {pages} {074201} (\bibinfo {year} {2009})}\BibitemShut
  {NoStop}%
\bibitem [{\citenamefont {Poole}\ \emph {et~al.}(1972-1973)\citenamefont
  {Poole}, \citenamefont {Leckey}, \citenamefont {Jenkin},\ and\ \citenamefont
  {Liesegang}}]{Poole:JESRP1:371}%
  \BibitemOpen
  \bibfield  {author} {\bibinfo {author} {\bibfnamefont {R.~T.}\ \bibnamefont
  {Poole}}, \bibinfo {author} {\bibfnamefont {R.~C.~G.}\ \bibnamefont
  {Leckey}}, \bibinfo {author} {\bibfnamefont {J.~G.}\ \bibnamefont {Jenkin}},
  \ and\ \bibinfo {author} {\bibfnamefont {J.}~\bibnamefont {Liesegang}},\
  }\bibfield  {title} {\enquote {\bibinfo {title} {Photoelectron angular
  distribution from gold},}\ }\href {\doibase 10.1016/0368-2048(72)80038-0}
  {\bibfield  {journal} {\bibinfo  {journal} {J.\ Elec.\ Spec.\ Rel.\ Phen.}\
  }\textbf {\bibinfo {volume} {2}},\ \bibinfo {pages} {371--376} (\bibinfo
  {year} {1972-1973})}\BibitemShut {NoStop}%
\bibitem [{\citenamefont {Pei}\ and\ \citenamefont
  {Berglund}(2002)}]{Pei:JJAP41:L52}%
  \BibitemOpen
  \bibfield  {author} {\bibinfo {author} {\bibfnamefont {Zeting}\ \bibnamefont
  {Pei}}\ and\ \bibinfo {author} {\bibfnamefont {C.~Neil}\ \bibnamefont
  {Berglund}},\ }\bibfield  {title} {\enquote {\bibinfo {title} {Angular
  distribution of photoemission from gold thin films},}\ }\href {\doibase  10.1143/JJAP.41.L52} {\bibfield  {journal} {\bibinfo  {journal} {Jpn.\ J.\
  Appl.\ Phys.}\ }\textbf {\bibinfo {volume} {41}},\ \bibinfo {pages}
  {L52--L54} (\bibinfo {year} {2002})}\BibitemShut {NoStop}%
\bibitem [{\citenamefont {Fann}\ \emph
  {et~al.}(1992{\natexlab{a}})\citenamefont {Fann}, \citenamefont {Storz},
  \citenamefont {Tom},\ and\ \citenamefont {Bokor}}]{Fann:PRL68:2834}%
  \BibitemOpen
  \bibfield  {author} {\bibinfo {author} {\bibfnamefont {W.~S.}\ \bibnamefont
  {Fann}}, \bibinfo {author} {\bibfnamefont {R.}~\bibnamefont {Storz}},
  \bibinfo {author} {\bibfnamefont {H.~W~K}\ \bibnamefont {Tom}}, \ and\
  \bibinfo {author} {\bibfnamefont {J.}~\bibnamefont {Bokor}},\ }\bibfield
  {title} {\enquote {\bibinfo {title} {{Direct measurement of nonequilibrium
  electron-energy distributions in sub-picosecond laser-heated gold films}},}\
  }\href {\doibase 10.1016/0039-6028(93)90985-S} {\bibfield  {journal}
  {\bibinfo  {journal} {Phys.\ Rev.\ Lett.}\ }\textbf {\bibinfo {volume}
  {68}},\ \bibinfo {pages} {2834--2837} (\bibinfo {year}
  {1992}{\natexlab{a}})}\BibitemShut {NoStop}%
\bibitem [{\citenamefont {Fann}\ \emph
  {et~al.}(1992{\natexlab{b}})\citenamefont {Fann}, \citenamefont {Storz},
  \citenamefont {Tom},\ and\ \citenamefont {Bokor}}]{Fann:PRB46:13592}%
  \BibitemOpen
  \bibfield  {author} {\bibinfo {author} {\bibfnamefont {W.~S.}\ \bibnamefont
  {Fann}}, \bibinfo {author} {\bibfnamefont {R.}~\bibnamefont {Storz}},
  \bibinfo {author} {\bibfnamefont {H.~W~K}\ \bibnamefont {Tom}}, \ and\
  \bibinfo {author} {\bibfnamefont {J.}~\bibnamefont {Bokor}},\ }\bibfield
  {title} {\enquote {\bibinfo {title} {{Electron thermalization in gold}},}\
  }\href {\doibase 10.1103/PhysRevB.46.13592} {\bibfield  {journal} {\bibinfo
  {journal} {Phys.\ Rev.\ B}\ }\textbf {\bibinfo {volume} {46}},\ \bibinfo
  {pages} {13592--13595} (\bibinfo {year} {1992}{\natexlab{b}})},\ \Eprint
  {http://arxiv.org/abs/1011.1669} {arXiv:1011.1669 [physics]} \BibitemShut
  {NoStop}%
\bibitem [{\citenamefont {Aeschlimann}\ \emph {et~al.}(1995)\citenamefont
  {Aeschlimann}, \citenamefont {Schmuttenmaer}, \citenamefont {Elsayed-Ali},
  \citenamefont {Miller}, \citenamefont {Cao}, \citenamefont {Gao},\ and\
  \citenamefont {Mantell}}]{Aeschlimann:JCP102:8606}%
  \BibitemOpen
  \bibfield  {author} {\bibinfo {author} {\bibfnamefont {M.}~\bibnamefont
  {Aeschlimann}}, \bibinfo {author} {\bibfnamefont {C.~A.}\ \bibnamefont
  {Schmuttenmaer}}, \bibinfo {author} {\bibfnamefont {H.~E.}\ \bibnamefont
  {Elsayed-Ali}}, \bibinfo {author} {\bibfnamefont {R.~J.~D.}\ \bibnamefont
  {Miller}}, \bibinfo {author} {\bibfnamefont {J.}~\bibnamefont {Cao}},
  \bibinfo {author} {\bibfnamefont {Y.}~\bibnamefont {Gao}}, \ and\ \bibinfo
  {author} {\bibfnamefont {D.~A.}\ \bibnamefont {Mantell}},\ }\bibfield
  {title} {\enquote {\bibinfo {title} {Observation of surface enhanced
  multiphoton photoemission from metal surfaces in the short pulse limit},}\
  }\href {\doibase 10.1063/1.468962} {\bibfield  {journal} {\bibinfo  {journal}
  {J.\ Chem.\ Phys.}\ }\textbf {\bibinfo {volume} {102}},\ \bibinfo {pages}
  {8606--8613} (\bibinfo {year} {1995})}\BibitemShut {NoStop}%
\bibitem [{\citenamefont {Banfi}\ \emph {et~al.}(2005)\citenamefont {Banfi},
  \citenamefont {Giannetti}, \citenamefont {Ferrini}, \citenamefont
  {Galimberti}, \citenamefont {Pagliara}, \citenamefont {Fausti},\ and\
  \citenamefont {Parmigiani}}]{Banfi:PRL94:037601}%
  \BibitemOpen
  \bibfield  {author} {\bibinfo {author} {\bibfnamefont {Francesco}\
  \bibnamefont {Banfi}}, \bibinfo {author} {\bibfnamefont {Claudio}\
  \bibnamefont {Giannetti}}, \bibinfo {author} {\bibfnamefont {Gabriele}\
  \bibnamefont {Ferrini}}, \bibinfo {author} {\bibfnamefont {Gianluca}\
  \bibnamefont {Galimberti}}, \bibinfo {author} {\bibfnamefont {Stefania}\
  \bibnamefont {Pagliara}}, \bibinfo {author} {\bibfnamefont {Daniele}\
  \bibnamefont {Fausti}}, \ and\ \bibinfo {author} {\bibfnamefont {Fulvio}\
  \bibnamefont {Parmigiani}},\ }\bibfield  {title} {\enquote {\bibinfo {title}
  {Experimental evidence of above-threshold photoemission in solids},}\ }\href
  {\doibase 10.1103/PhysRevLett.94.037601} {\bibfield  {journal} {\bibinfo
  {journal} {Phys.\ Rev.\ Lett.}\ }\textbf {\bibinfo {volume} {94}},\ \bibinfo
  {pages} {037601} (\bibinfo {year} {2005})},\ \Eprint
  {http://arxiv.org/abs/1201.3049} {arXiv:1201.3049 [physics]} \BibitemShut
  {NoStop}%
\bibitem [{\citenamefont {Corkum}(1993)}]{Corkum:PRL71:1994}%
  \BibitemOpen
  \bibfield  {author} {\bibinfo {author} {\bibfnamefont {P.~B.}\ \bibnamefont
  {Corkum}},\ }\bibfield  {title} {\enquote {\bibinfo {title} {Plasma
  perspective on strong-field multiphoton ionization},}\ }\href {\doibase  10.1103/PhysRevLett.71.1994} {\bibfield  {journal} {\bibinfo  {journal}
  {Phys.\ Rev.\ Lett.}\ }\textbf {\bibinfo {volume} {71}},\ \bibinfo {pages}
  {1994--1997} (\bibinfo {year} {1993})}\BibitemShut {NoStop}%
\end{thebibliography}%


\begin{thebibliography}{6}%
\makeatletter
\providecommand \@ifxundefined [1]{%
 \@ifx{#1\undefined}
}%
\providecommand \@ifnum [1]{%
 \ifnum #1\expandafter \@firstoftwo
 \else \expandafter \@secondoftwo
 \fi
}%
\providecommand \@ifx [1]{%
 \ifx #1\expandafter \@firstoftwo
 \else \expandafter \@secondoftwo
 \fi
}%
\providecommand \natexlab [1]{#1}%
\providecommand \enquote  [1]{``#1''}%
\providecommand \bibnamefont  [1]{#1}%
\providecommand \bibfnamefont [1]{#1}%
\providecommand \citenamefont [1]{#1}%
\providecommand \href@noop [0]{\@secondoftwo}%
\providecommand \href [0]{\begingroup \@sanitize@url \@href}%
\providecommand \@href[1]{\@@startlink{#1}\@@href}%
\providecommand \@@href[1]{\endgroup#1\@@endlink}%
\providecommand \@sanitize@url [0]{\catcode `\\12\catcode `\$12\catcode
  `\&12\catcode `\#12\catcode `\^12\catcode `\_12\catcode `\%12\relax}%
\providecommand \@@startlink[1]{}%
\providecommand \@@endlink[0]{}%
\providecommand \url  [0]{\begingroup\@sanitize@url \@url }%
\providecommand \@url [1]{\endgroup\@href {#1}{\urlprefix }}%
\providecommand \urlprefix  [0]{URL }%
\providecommand \Eprint [0]{\href }%
\providecommand \doibase [0]{http://dx.doi.org/}%
\providecommand \selectlanguage [0]{\@gobble}%
\providecommand \bibinfo  [0]{\@secondoftwo}%
\providecommand \bibfield  [0]{\@secondoftwo}%
\providecommand \translation [1]{[#1]}%
\providecommand \BibitemOpen [0]{}%
\providecommand \bibitemStop [0]{}%
\providecommand \bibitemNoStop [0]{.\EOS\space}%
\providecommand \EOS [0]{\spacefactor3000\relax}%
\providecommand \BibitemShut  [1]{\csname bibitem#1\endcsname}%
\let\auto@bib@innerbib\@empty
\bibitem [{\citenamefont {{Scientific Instrument Services Inc.,
  USA}}(2011)}]{Simion:8.1}%
  \BibitemOpen
  \bibfield  {author} {\bibinfo {author} {\bibnamefont {{Scientific Instrument
  Services Inc., USA}}},\ }\href@noop {} {\enquote {\bibinfo {title} {Simion
  8.1},}\ } (\bibinfo {year} {2011}),\ \bibinfo {note}
  {{URL}:~\url{http://simion.com}}\BibitemShut {NoStop}%
\bibitem [{\citenamefont {M{\"u}ller}\ \emph {et~al.}(2015)\citenamefont
  {M{\"u}ller}, \citenamefont {Trippel}, \citenamefont {D{\l}ugo{\l}\k{e}cki},\
  and\ \citenamefont {K{\"u}pper}}]{Mueller:JPB48:244001}%
  \BibitemOpen
  \bibfield  {author} {\bibinfo {author} {\bibfnamefont {Nele L.~M.}\
  \bibnamefont {M{\"u}ller}}, \bibinfo {author} {\bibfnamefont {Sebastian}\
  \bibnamefont {Trippel}}, \bibinfo {author} {\bibfnamefont {Karol}\
  \bibnamefont {D{\l}ugo{\l}\k{e}cki}}, \ and\ \bibinfo {author} {\bibfnamefont
  {Jochen}\ \bibnamefont {K{\"u}pper}},\ }\bibfield  {title} {\enquote
  {\bibinfo {title} {Electron gun for diffraction experiments on controlled
  molecules},}\ }\href {\doibase 10.1088/0953-4075/48/24/244001} {\bibfield
  {journal} {\bibinfo  {journal} {J.\ Phys.\ B}\ }\textbf {\bibinfo {volume}
  {48}},\ \bibinfo {pages} {244001} (\bibinfo {year} {2015})},\ \Eprint
  {http://arxiv.org/abs/1507.02530} {arXiv:1507.02530 [physics]} \BibitemShut
  {NoStop}%
\bibitem [{\citenamefont {Poole}\ \emph {et~al.}(1972-1973)\citenamefont
  {Poole}, \citenamefont {Leckey}, \citenamefont {Jenkin},\ and\ \citenamefont
  {Liesegang}}]{Poole:JESRP1:371}%
  \BibitemOpen
  \bibfield  {author} {\bibinfo {author} {\bibfnamefont {R.~T.}\ \bibnamefont
  {Poole}}, \bibinfo {author} {\bibfnamefont {R.~C.~G.}\ \bibnamefont
  {Leckey}}, \bibinfo {author} {\bibfnamefont {J.~G.}\ \bibnamefont {Jenkin}},
  \ and\ \bibinfo {author} {\bibfnamefont {J.}~\bibnamefont {Liesegang}},\
  }\bibfield  {title} {\enquote {\bibinfo {title} {Photoelectron angular
  distribution from gold},}\ }\href {\doibase 10.1016/0368-2048(72)80038-0}
  {\bibfield  {journal} {\bibinfo  {journal} {J.\ Elec.\ Spec.\ Rel.\ Phen.}\
  }\textbf {\bibinfo {volume} {2}},\ \bibinfo {pages} {371--376} (\bibinfo
  {year} {1972-1973})}\BibitemShut {NoStop}%
\bibitem [{\citenamefont {Pei}\ and\ \citenamefont
  {Berglund}(2002)}]{Pei:JJAP41:L52}%
  \BibitemOpen
  \bibfield  {author} {\bibinfo {author} {\bibfnamefont {Zeting}\ \bibnamefont
  {Pei}}\ and\ \bibinfo {author} {\bibfnamefont {C.~Neil}\ \bibnamefont
  {Berglund}},\ }\bibfield  {title} {\enquote {\bibinfo {title} {Angular
  distribution of photoemission from gold thin films},}\ }\href {\doibase  10.1143/JJAP.41.L52} {\bibfield  {journal} {\bibinfo  {journal} {Jpn.\ J.\
  Appl.\ Phys.}\ }\textbf {\bibinfo {volume} {41}},\ \bibinfo {pages}
  {L52--L54} (\bibinfo {year} {2002})}\BibitemShut {NoStop}%
\bibitem [{\citenamefont {Berglund}\ and\ \citenamefont
  {Spicer}(1964)}]{Berglund:PR136:A1044}%
  \BibitemOpen
  \bibfield  {author} {\bibinfo {author} {\bibfnamefont {C.~N.}\ \bibnamefont
  {Berglund}}\ and\ \bibinfo {author} {\bibfnamefont {W.~E.}\ \bibnamefont
  {Spicer}},\ }\bibfield  {title} {\enquote {\bibinfo {title} {Photoemission
  studies of copper and silver: Experiment},}\ }\href {\doibase  10.1103/PhysRev.136.A1044} {\bibfield  {journal} {\bibinfo  {journal} {Phys.\
  Rev.}\ }\textbf {\bibinfo {volume} {136}},\ \bibinfo {pages} {A1044--A1064}
  (\bibinfo {year} {1964})}\BibitemShut {NoStop}%
\bibitem [{\citenamefont {Dasch}(1992)}]{Dasch:AO31:1146}%
  \BibitemOpen
  \bibfield  {author} {\bibinfo {author} {\bibfnamefont {Cameron~J.}\
  \bibnamefont {Dasch}},\ }\bibfield  {title} {\enquote {\bibinfo {title}
  {{One-dimensional tomography: a comparison of Abel, onion-peeling, and
  filtered backprojection methods}},}\ }\href {\doibase 10.1364/AO.31.001146}
  {\bibfield  {journal} {\bibinfo  {journal} {Applied Optics}\ }\textbf
  {\bibinfo {volume} {31}},\ \bibinfo {pages} {1146} (\bibinfo {year}
  {1992})}\BibitemShut {NoStop}%
\end{thebibliography}%
\end{document}


\title{Supplementary information: Velocity-map imaging for emittance characterization \mbox{of
      multiphoton-emitted electrons from a gold surface}}
\author{Hong Ye}\cfeldesy\uhhphys
\author{Sebastian Trippel}\email[Email:~]{sebastian.trippel@cfel.de}\cfeldesy\uhhcui
\author{Michele Di Fraia}\elettra\cfeldesy\uhhcui
\author{Arya Fallahi}\cfeldesy
\author{Oliver D.~Mücke}\cfeldesy\uhhcui
\author{Franz X.~Kärtner}\cfeldesy\uhhphys\uhhcui
\author{Jochen Küpper}\cfeldesy\uhhphys\uhhcui
\date{\today}
\maketitle

\section{Spectrometer characterization}
\begin{figure}[b]
   \centering
   \includegraphics[width=\linewidth]{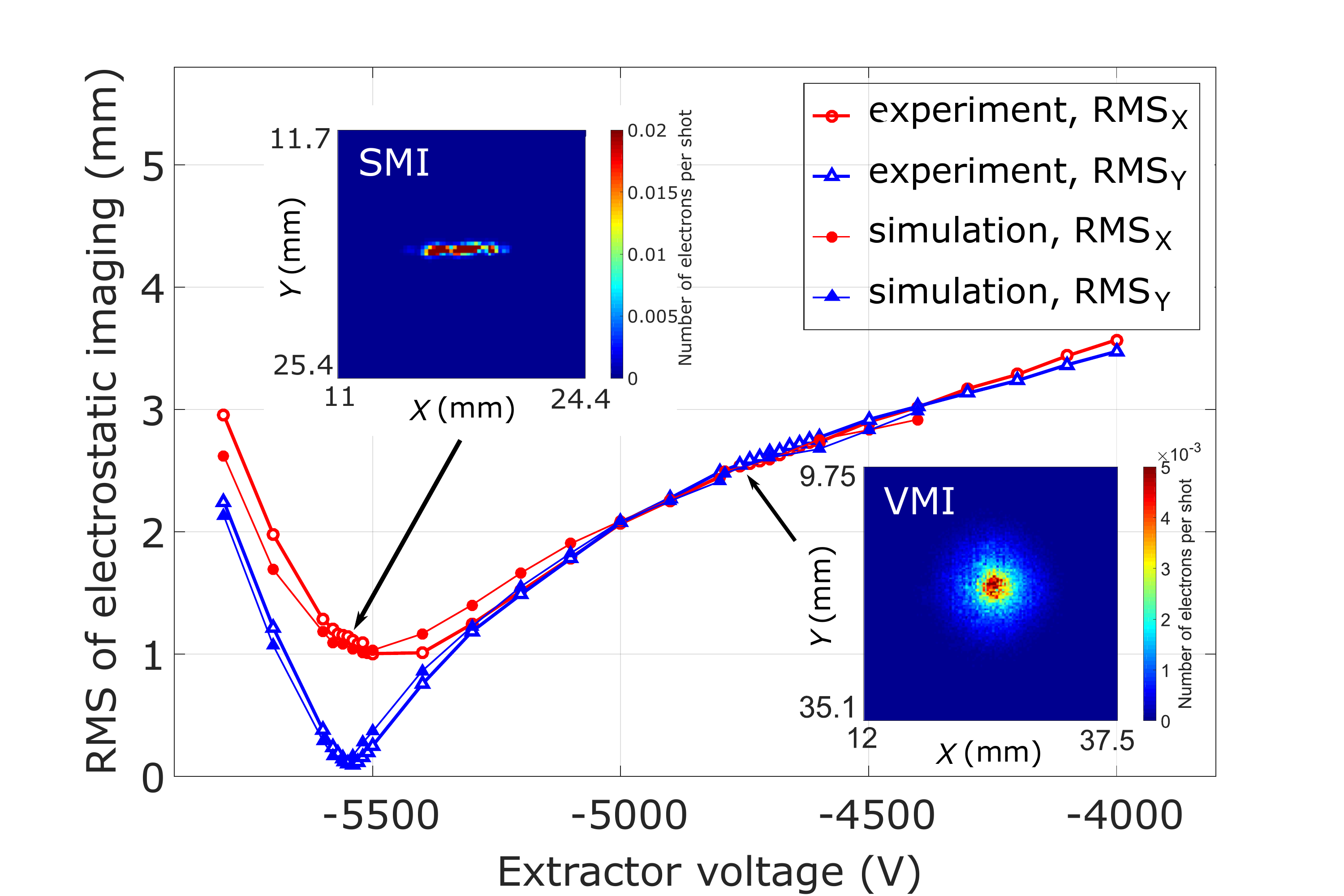}
   \caption{Experimental (hollow) and simulated (solid) root-mean-square deviations of electron
      spatial distributions on the 2D detector versus focusing extractor voltage in both $X$ and
      $Y$-directions. The insets show SMI and VMI detector images for the indicated positions.}
   \label{fig:extractor_scan}
\end{figure}

The electron spectrometer has been characterized experimentally, accompanied by simulations, in
order to determine the focusing conditions for the SMI and VMI modes; see \mainfig{1} for the
experimental setup. \autoref{fig:extractor_scan} shows the measured root-mean-square (RMS) in the
$X$- and $Y$-directions of the spatial electron distribution on the detector as a function of the
extractor voltage, together with the results from SIMION~\cite{Simion:8.1} electric field and
particle trajectory simulations. A similar behavior as in Ref.~\onlinecite{Mueller:JPB48:244001} is
observed.

The strongest focusing of the electron bunch onto the detector is achieved at an extractor potential
of -5560~V, which is thus identified as the SMI voltage. The RMS at this voltage shows the magnified
laser-surface-interaction area. The slightly different focusing behavior of the electron bunch in
the $X$ and $Y$-directions is attributed to the asymmetric initial electron bunch size, due to the
glancing incidence irradiation, and the finite kinetic energy of the electrons.

When increasing the extractor voltage, the electron bunch diverges. Based on our simulations, the
extractor voltage for VMI is approximately -4790~V. For a full calibration of the spectrometer, the
simulations were used to study the field configuration and the electron trajectories in those fields
for the given electrode configurations and the particles initial distributions. In
\autoref{fig:extractor_scan} the simulated RMS of the electron bunch, with electrostatic imaging, at
the detector position is plotted as function of extractor voltage. The simulations were carried out
given an initial spatial 2D Gaussian distribution of 2000 electrons for each simulated point. The
center of mass~(COM) of this distribution was given by $(X,Y)=(0,0)$ and a $Z$-coordinate matching the
sample surface with standard deviations of $\sigma_X=140~\um$ and $\sigma_Y=15~\um$. The initial
momentum distribution was given by a uniform half sphere with an uniform kinetic energy distribution
of electrons in the range of $[0.1,0.6]$~eV.

The COM of the electron distribution as a function of the initial starting position of the
electrons, \ie, the laser focus position on the sample, was used to experimentally calibrate the
voltage for velocity-map imaging.
\begin{figure}
   \centering
   \includegraphics[width=\linewidth]{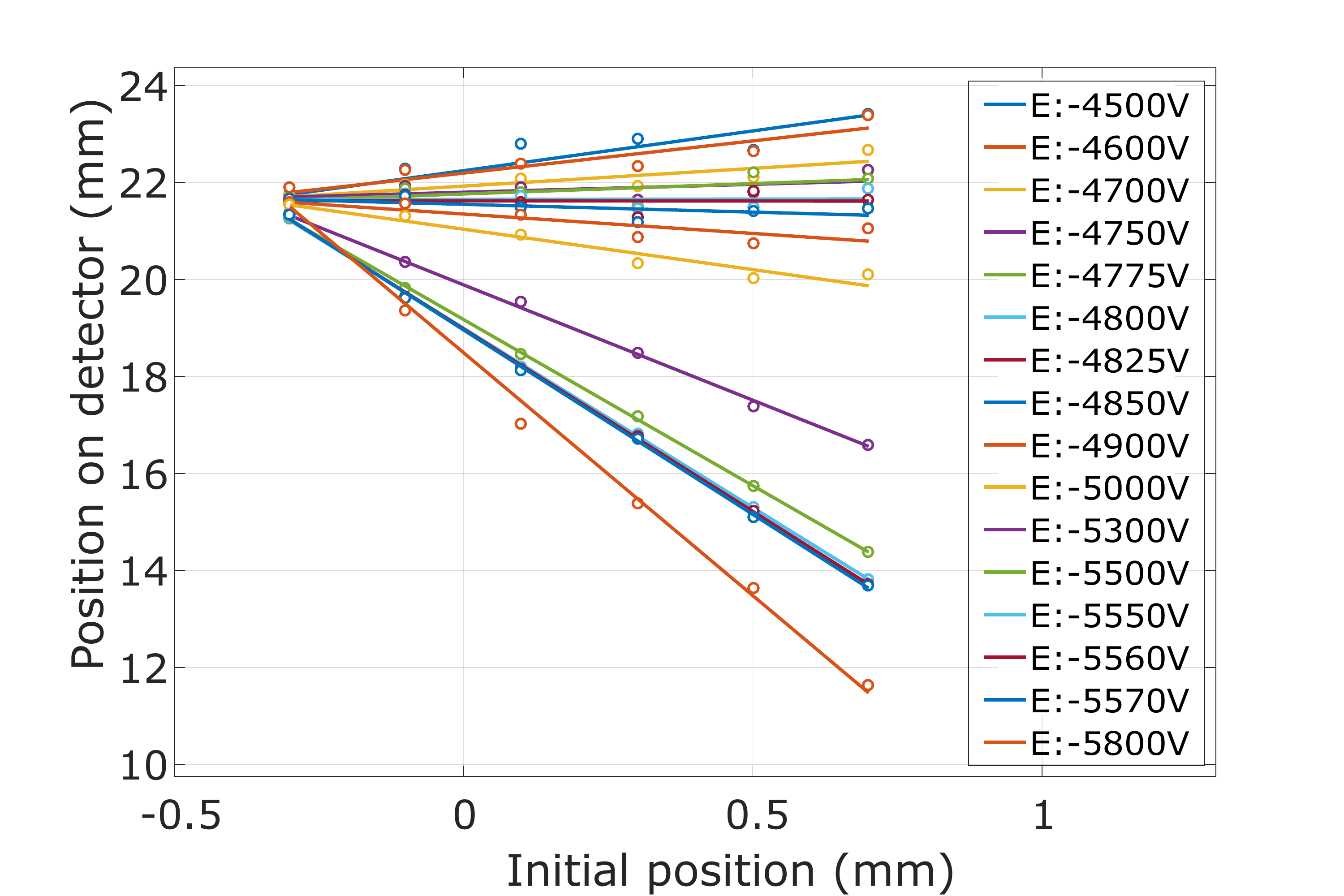}
   \caption{Position dependence of the center of mass of the electrostatic imaging on the detector
      on the initial source position for various extractor potentials from -5800~V to -4500~V.}
   \label{fig:VMI_mode}
\end{figure}
\autoref{fig:VMI_mode} shows the COM as function of the laser position for various voltages together
with straight-line fits. A decrease of the slope with decreasing extractor voltage is observed.
\begin{figure}
   \centering
   \includegraphics[width=\linewidth]{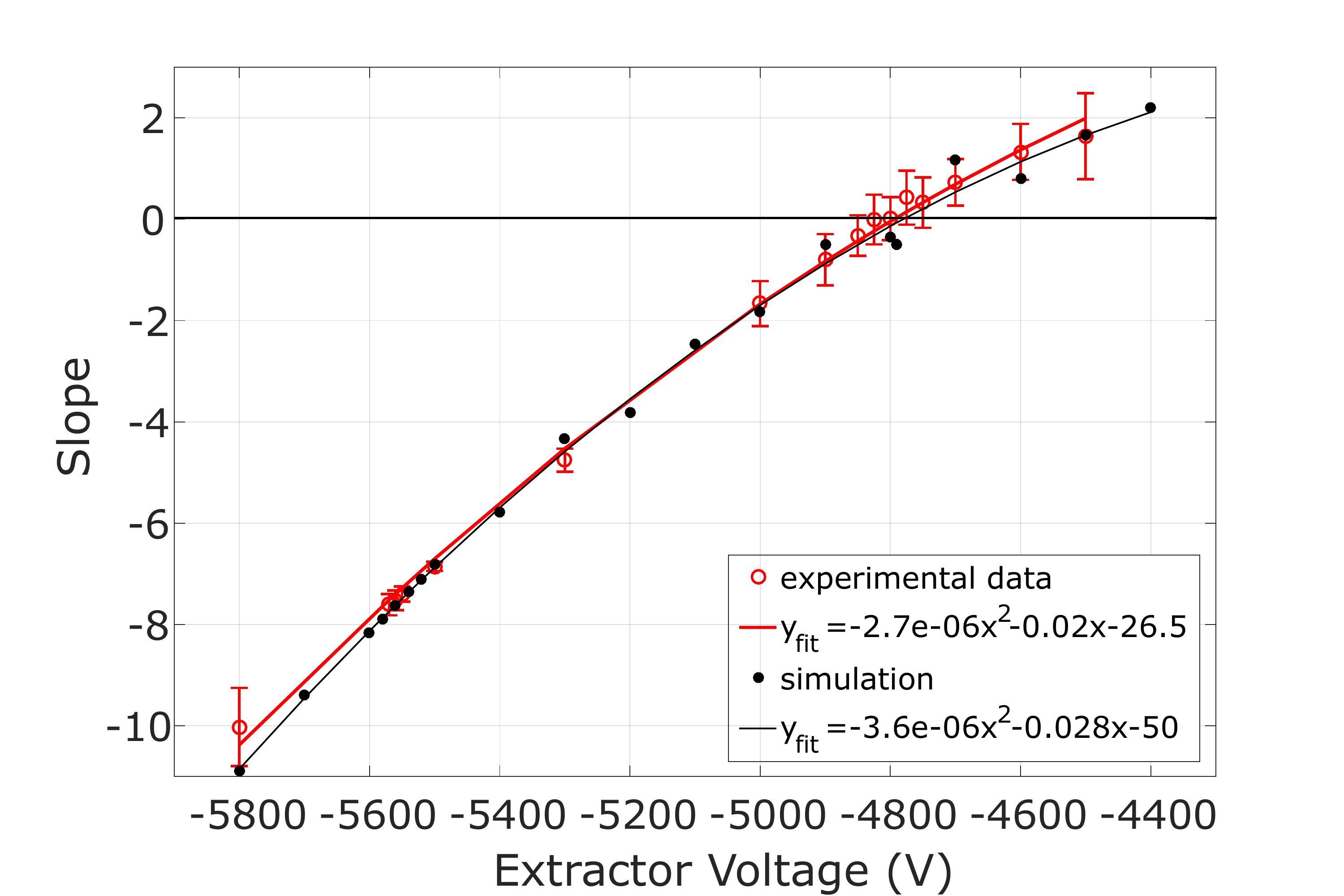}
   \caption{Slope of the experimental laser position dependent COM of the spatial distribution at
     the detector as function of the extractor voltage (red circles) with a quadratic fit (red
     line). Black points and the black line indicate the corresponding simulated results.}
   \label{fig:VMI_mode_slope}
\end{figure}
\autoref{fig:VMI_mode_slope} depicts the slope of each measurement in \autoref{fig:VMI_mode} as
function of the extractor voltage together with a quadratic fit and corresponding simulation
results. The error bar for the experimental points is given by the first-order coefficient error of
each fitting curve with 95~\% confidence bounds. VMI mode is obtained at the zero crossing of this
curve, \ie, at -4790~V, as for this voltage the distribution, to first order, becomes independent of
the starting position. The data shows a good agreement with the simulations, confirming that the
extractor voltage for operating the VMI is -4790~V. From the simulations the imaging setup is
calibrated regarding the transverse electron velocities to 8014~m/s/pixel on the detector.
\begin{table}
   \begin{center}
      \caption{Voltages (in V) applied for operation in SMI and VMI mode}
      \begin{tabular}{c@{\hspace{1em}}c@{\hspace{1em}}c@{\hspace{1em}}c@{\hspace{1em}}c}
        \hline \hline
        & Repeller & Extractor & Ground & Sample \\
        \hline
        SMI & -6000 & -5560 & 0 & -6000 \\
        VMI & -6000 & -4790 & 0 & -6000 \\
        \hline \hline
      \end{tabular}
   \end{center}
   \label{tab:SMI_VMI_voltage}
\end{table}
The resulting voltages for operation in the SMI and VMI modes are listed in
\autoref{tab:SMI_VMI_voltage}.

\begin{figure}
   \centering
   \includegraphics[width=\linewidth]{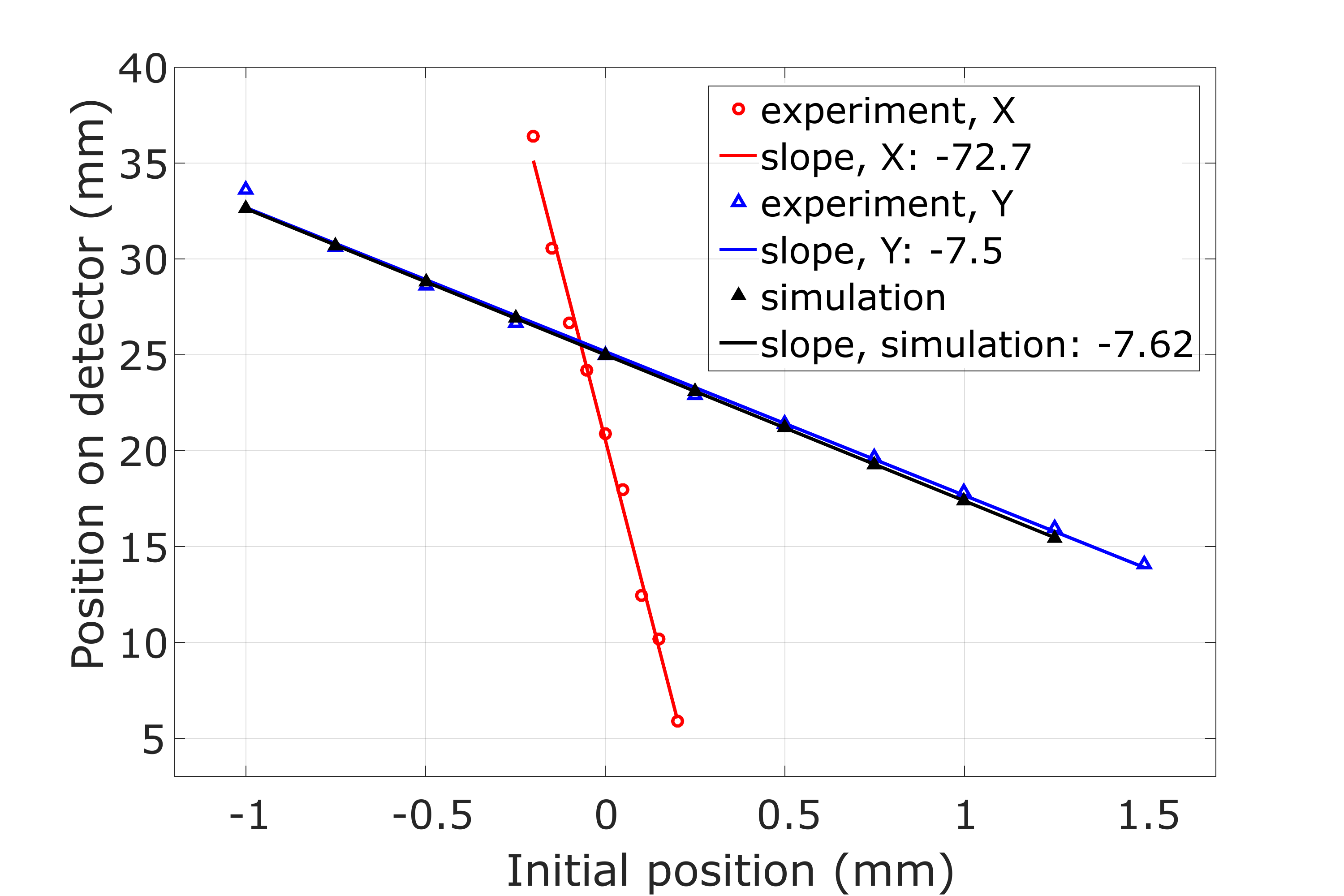}
   \caption{The COM of electrostatic imaging on the detector as a function of the initial source
      position for SMI mode, \ie, an extractor voltage of -5560~V. The slope in $Y$-direction is the
      spatial magnification factor. The ratio between $X$- and $Y$-directions confirms the incidence
      angle of the laser beam of \degree{84}.}
   \label{fig:SMI_mode}
\end{figure}
\autoref{fig:SMI_mode} shows experimental and theoretical COM of the electron distribution at the
detector as function of the lens position, that is used to focus the laser beam onto the sample for
SMI (E:-5560~V). The straight lines are fits to the data. The difference in the slope between the
$X$- and $Y$-directions is due to the glancing incidence angle $\theta$. The laser spot position on
sample moves $1/\cos\theta$ times farther in $X$ than in $Y$ when displacing the laser beam the same
distance by a translation stage. For the $Y$-direction we obtain a magnification factor of $\sim\!7.5$
from the fit. For the $X$-direction a slope of $\ordsim72.7$ is obtained. This results in a ratio of
9.7 between the two slopes that corresponds to an incident angle of $\degree{84}$. The SIMION
simulation results, also shown in \autoref{fig:SMI_mode}, are in good agreement with the data.

The focusing conditions for the SMI and VMI mode depend strongly on the position of the sample
inside the velocity-map imaging spectrometer.
\begin{figure}
   \centering
   \includegraphics[width=\linewidth]{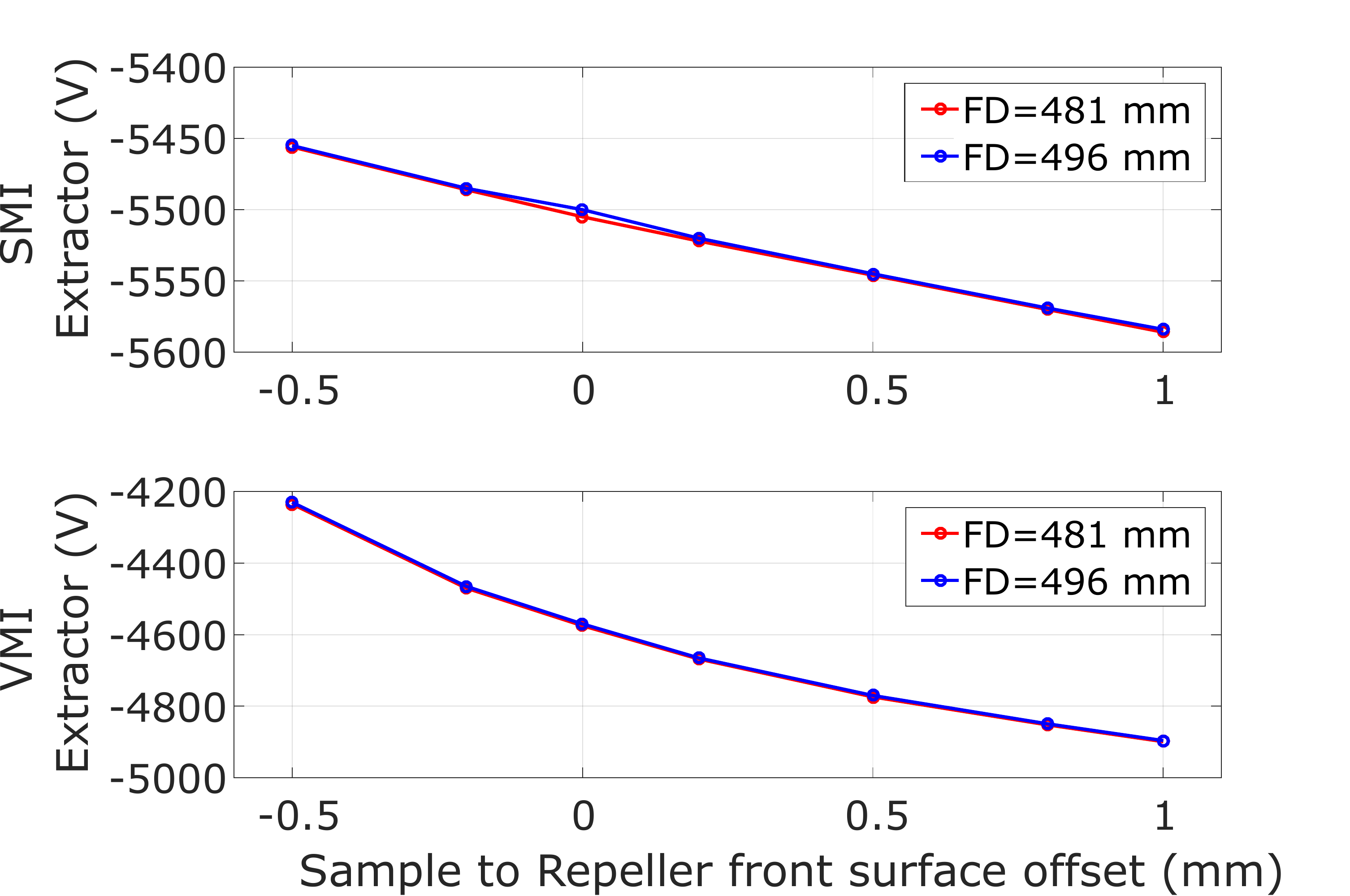}
   \caption{Top:~SMI and bottom:~VMI extractor potential for different position offset from sample
      front to the repeller front surfaces. FD stands for flying distance in the figure legend.}
   \label{fig:SMI_VMI_sample_position}
\end{figure}
\autoref{fig:SMI_VMI_sample_position} shows the simulated extractor voltages necessary for SMI and
VMI mode for various sample displacements with respect to the front surface of the repeller plate.
These simulations show that either the sample position has to be known, or at least be reproduced,
to a very high precision or calibration measurements have to be performed when a new sample is
inserted into the spectrometer. Fortunately, with the protocol described in our manuscript this
calibration can be done quickly. In addition the dependence of the extractor voltage on the flight
distance has been investigated (red points and lines). Our simulations show that this uncertainty is
uncritical compared to the exact sample position in the spectrometer.

\section{Reconstruction algorithm}

Our reconstruction algorithm for the conversion of the 2D projected velocity distribution to the 3D
distribution is based on the assumption that the angular distribution of the photoemitted electrons
is known. For our simulations, a cosine function
$I(\theta)\propto\cos\theta$~\cite{Poole:JESRP1:371, Pei:JJAP41:L52}, derived from the
Berglund-Spicer model~\cite{Berglund:PR136:A1044} as discussed in the main text, is applied in the
algorithm. In addition, it is assumed that for multiphoton emission the angular distribution is
independent of the modulus of the three dimensional velocity vector. The 3D velocity distribution is
then obtained from the 2D projected distribution by a matrix method similar to Onion
Peeling~\cite{Dasch:AO31:1146}. For multiphoton emission from a planar Au surface, the electrons are
assumed to be photoemitted within a half sphere of $\varphi\in[0,~2\pi]$, $\theta\in[0,~\pi/2]$. The
photoemitted electron distribution has cylindrical symmetry with respect to the surface normal of
the sample.

\begin{figure}[h]
	\centering
	\includegraphics[width=\linewidth]{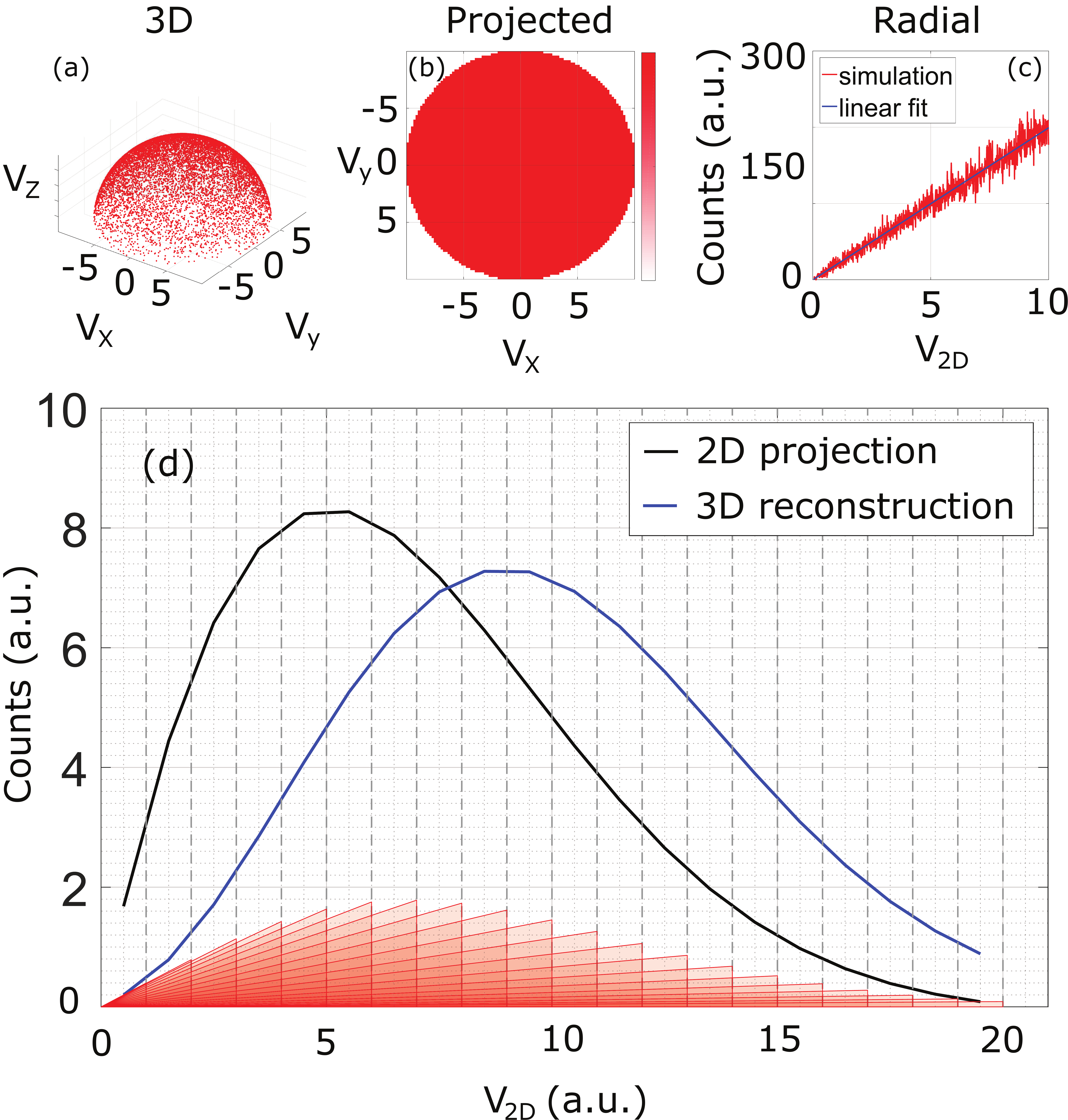}
	\caption{(a-c) Representation of a simulated electron bunch with a single 3D velocity $v_i$ and an
		angular distribution of a cosine function: (a) in 3D, forming a spherical surface; (b) in 2D,
		yielding a uniform distribution in the detector plane; (c) in 1D, showing a linearly increasing
		radial velocity $v_{2D}$ with distance from distribution COM. (d) A conceptual diagram of the reconstruction
		algorithm: The area of each red triangle at the bottom indicates the number of photoemitted
		electrons having the same 3D velocity. The corresponding distribution curve is plotted as blue
		curve. The black curve is the 2D projection distribution curve, summing up the number of
		photoemitted electrons within each interval of the same transverse velocity. The gray dashed lines
		indicate the transverse-velocity intervals used in this projection.}
	\label{fig:concept_diagram}
\end{figure}

\subautoref{fig:concept_diagram}{a} shows a scatter plot for a single 3D velocity $v_i$ distribution
given by $f(v,\theta)=\delta(v-v_i)\cos\theta$. \subautoref{fig:concept_diagram}{b} shows the
projection of this distribution onto the 2D detector surface. It can be derived that the projected
velocity distribution for this special case is
\begin{equation*}
   P_i(v_x,v_y) = \int f(v,\theta) \dif{v_z} = \begin{cases}
      C & \text{for~} v_{x,y}<v_i, \\
      0 & \text{otherwise}
   \end{cases}
\end{equation*}
where $C$ is a constant. As shown in \subautoref{fig:concept_diagram}{b}, the projected velocity
distribution of $f(v,\theta)$ is constant inside the circular phase-space area of radius $v_i$.
Furthermore, \subautoref{fig:concept_diagram}{c} shows the radial distribution obtained from the
projected velocity distribution given by
\begin{equation*}
   \rho_i(v_\text{2D}) = \int P_i(v_x,v_y) \dif{\theta_\text{2D}} = \begin{cases}
      2\pi C \cdot v_\text{2D} & \text{for~} v_{2D}<v_i\\
      0 & \text{otherwise}
   \end{cases}
\end{equation*}
where $v_{2D}=\sqrt{{v_x}^2+{v_y}^2}$. In the reconstruction, each radial distribution
$\rho_i(v_{2D})$ is built up by a triangle as sketched in \subautoref{fig:concept_diagram}{d}. $v_i$
is taken equally spaced and form the intervals confined by the neighboring gray dashed lines. The 2D
projected distribution is related to the 3D distribution $f_i$ by a transfer matrix $\mathbf{M}$.
\begin{equation}
   \rho_i = \mathbf{M}f_i,
\end{equation}
with $\mathbf{M}$ given by:
\begin{equation}
   \mathbf{M}=
   \begin{pmatrix} 
      1 & 1/4 & 1/9 & 1/16 & \cdots \\
      0 & 3/4 & 3/9 & 3/16 & \cdots \\
      0 & 0 & 5/9 & 5/16 & \cdots \\
      0 & 0 & 0 & 7/16 & \cdots \\
      \vdots & \vdots & \vdots & \vdots & \ddots \\
   \end{pmatrix}
\end{equation}

The 3D distribution can finally be obtained by inversion of the measured 2D-projected distribution
\begin{equation}
   f_i=\mathbf{M}^{-1}\rho_i.
\end{equation}

\bibliography{string,cmi}